\documentclass[usegraphicx,usenatbib,a4paper]{mn2e}
\usepackage{aas_macros,graphicx,times,multirow}
\usepackage{amssymb}
\usepackage{graphicx,natbib}
\usepackage{lscape}
\usepackage{float}
\usepackage[T1]{fontenc}
\usepackage{aecompl}
\title[New magnetic CVs observed with \XMM]
{Broad-band characteristics of seven new hard X-ray selected cataclysmic variables}
\author[F. Bernardini et al.]                                                    
{F.~Bernardini,$^{1,2}$\thanks{E-mail:bernardini@nyu.edu} 
D.~de Martino,$^{2}$ 
K.~Mukai,$^{3,4}$
D. M.~Russell,$^{1}$
M.~Falanga,$^{5}$
N.~Masetti,$^{6,7}$
\newauthor
C.~Ferrigno,$^{8}$
G.~Israel$^{6}$
\\
$^1$ New York University Abu Dhabi, Saadiyat Island, Abu Dhabi, 129188, United Arab Emirates\\
$^2$ INAF $-$ Osservatorio Astronomico di Capodimonte, Salita Moiariello 16, I-80131 Napoli, Italy\\
$^3$ CRESST and X-Ray Astrophysics Laboratory, NASA Goddard Space Flight Center, Greenbelt, MD 20771, USA\\
$^4$ Department of Physics, University of Maryland, Baltimore County, 1000 Hilltop Circle, Baltimore, MD 21250, USA\\
$^5$ International Space Science Institute (ISSI), Hallerstrasse 6, CH-3012 Bern, Switzerland\\
$^6$ INAF - Istituto di Astrofisica Spaziale e Fisica Cosmica, Via Gobetti 101, I-40129, Bologna, Italy \\
$^7$ Departamento de Ciencias F{\'i}sicas, Universidad Andr{\'e}s Bello, Fern{\'a}ndez Concha 700, Las Condes, Santiago, Chile\\
$^8$ ISDC, Department of astronomy, University of Geneva, chemin d'\'Ecogia, 16 CH-1290 Versoix, Switzerland \\
$^9$ INAF - Osservatorio Astronomico di Roma, via Frascati 33, I-00040 Monteporzio Catone, Roma, Italy}

\date{}
\pagerange{\pageref{firstpage}--\pageref{lastpage}}
\def\Swift{{\em Swift}}

\def\XMM{{\em XMM-Newton}}

\def\ergscm{$\rm erg\,cm^{-2}\,s^{-1}$}
\def\INT{{\em INTEGRAL}\,}

\def\aa092{Swift\,J0927.7-6945}
\def\bb095{Swift\,J0958.0-4208}
\def\cc170{Swift\,J1701.3-4304}
\def\dd070{Swift\,J0706.8+0325}
\def\ee211{Swift\,J2113.5+5422}
\def\ff080{PBC\,J0801.2-4625}
\def\gg074{Swift\,J0746.3-1608}

\begin{document}

\label{firstpage}

\maketitle

\begin{abstract}

We present timing and spectral analysis of a sample of seven  
hard X-ray selected Cataclysmic Variable candidates based on simultaneous X-ray and optical observations collected with \XMM , complemented with \Swift/BAT and \INT/IBIS hard X-ray data and ground-based optical photometry. 
For six sources, X-ray pulsations are detected for the first time 
in the range $\rm \sim296-6098\,s$, identifying them as members of
the magnetic class. Swift\,J0927.7-6945, Swift\,J0958.0-4208, Swift\,J1701.3-4304, 
Swift\,J2113.5+5422, and possibly PBC\,J0801.2-4625, are 
Intermediate Polars (IPs), while 
Swift\,J0706.8+0325 is a short (1.7\,h) orbital period Polar, the 11$^{\rm th}$ hard 
X-ray selected identified so far. X-ray orbital modulation is also observed in Swift\,J0927.7-6945 (5.2\,h) and Swift\,J2113.5+5422 (4.1\,h).  
Swift\,J1701.3-4304 is discovered as the longest orbital period (12.8\,h) 
deep eclipsing IP. The spectra of the magnetic systems reveal
optically thin multi-temperature emission between 0.2 and 60 keV. 
Energy dependent spin pulses and the orbital modulation in Swift\,J0927.7-6945 and Swift\,J2113.5+5422 are due to intervening local high density 
absorbing material  ($\rm N_H\sim10^{22-23}\,cm^{-2}$). In Swift\,J0958.0-4208 and Swift\,J1701.3-4304, a soft X-ray blackbody (kT$\sim$50 and $\sim$80 eV) is detected, adding them to the growing group of "soft" IPs. White dwarf masses are determined in the range $\rm \sim0.58-1.18\,M_{\odot}$, indicating massive accreting primaries in five of them. Most sources accrete at rates lower than the expected secular value for their orbital period. 
Formerly proposed as a long-period (9.4\,h) novalike CV, 
Swift\,J0746.3-1608 shows peculiar spectrum and light curves suggesting
either an atypical low-luminosity CV or a low mass X-ray binary.
\end{abstract}


\section{Introduction}

Cataclysmic variables (CVs) are close binary stellar systems where a compact 
primary, a white dwarf (WD), 
accretes matter from a low mass (${\rm M}\lesssim1\,{\rm M}_{\odot}$) Roche lobe 
filling main-sequence or subgiant secondary. 
About one quarter of the whole CV class host strongly magnetized WDs 
($10^{5}<{\rm B}<2.4\times10^{8}$ G). 
The magnetic systems (MCVs) are further divided in two subclasses, 
depending on the WD magnetic field intensity and degree of synchronism 
\citep[see e.g.][]{Cropper90,Warner95,hellier14,ferrario15,mukai17}.
Those showing intense optical/near-infrared (nIR) polarised emission, 
the so-called polars (or AM Her stars), 
have stronger magnetic fields (B$\gtrsim10^{7}$ G) 
which synchronize the WD rotation with the orbital period 
(${\rm P}_{spin=\omega} \sim {\rm P}_{orb=\Omega} \sim$ 80\,m -- few hours). 
The IPs (also known as DQ Her stars) instead are 
asynchronous systems (${\rm P}_{\omega}/{\rm P}_{\Omega}<1$), with a few exceptions do not show detectable polarized optical/nIR  
emission, and are consequently believed to host WDs with lower magnetic 
field strengths (B$\leq10^{6}$ G). 
Both groups are characterised by periodic X-ray emission at the WD 
rotational period, a clear signature of magnetic accretion.
IPs mainly populate the orbital period distribution above the so-called  
2--3 h orbital period gap, while polars are generally found
below it \citep{Warner95}, suggesting that IPs may eventually 
evolve into polars \citep{norton04}.
\\
The way in which the material lost from the companion is accreted by the WD, 
mainly depends on the WD magnetic field intensity. Eventually, when the flow gets 
closer to the WD, it is magnetically channeled along the field lines. Due to the
strong dipolar field in polars, matter lost by the companion is accreted 
directly from a stream, while IPs may accrete from a stream or 
through a disc, depending on the magnetic moment and the degree of 
asynchronysm \citep{norton04,norton08}. In the latter case accretion proceeds
from the disc to the WD surface on arc-shaped accretion curtains \citep[][]{rosen88}. 
Disc-overflow (a hybrid accretion mode), may also occur depending
on the mass accretion rate \citep[][]{hellier95,norton97}.

Since the velocities in the accretion flow are supersonic a stand-off shock forms. 
The temperature in the post shock region is high ($\sim10-80$ keV) and 
the flow cools and slows down via bremsstrahlung (hard X-ray) and cyclotron 
(optical/nIR) radiation \citep{aizu73,wu94,cropper99}. The efficiency 
of the cooling mechanisms depends on the WD magnetic field intensity and the
local mass accretion rate \citep{woelk96,fischer01}. 
The hard X-rays and cyclotron emission are partially absorbed and reflected from the WD surface. The
thermalised emission from the WD pole gives rise to a non-negligible soft 
($\sim20-60$ eV) X-ray blackbody component, formerly believed a 
characteristic of the polar systems only \citep{beuermann99}. The presence of this
component has been challenged by \XMM\ observations showing an  
increasing number of polars without it \citep[see e.g.][]{ramsay04b,ramsay09,bernardini14}. On the other hand, 
while {\it ROSAT} initially
detected only a handful of "soft" IP systems \citep{haberl_motch95}, 
\XMM\ has later revealed that a soft X-ray blackbody is also present in a 
non-negligible number of IPs
\citep{evans_hellier07,demartino04,anzolin08,
demartino08,bernardini12}.
Compton reflection from the WD surface, considered nearly neutral, 
produces a non-negligible Fe $K_{\alpha}$ line at 6.4 keV as observed
in all MCVs \citep{EzukaIshida99}, as well as a Compton reflection continuum
peaking at energies above 10 keV 
\citep{done_and_magdziarz98}, which was unambiguously 
detected in recent joint \XMM\ and {\it NuSTAR} observations of 
three bright IPs \citep{mukai15}. 
Complex absorption from neutral material in the dense 
($\rm N_H$ up to $\rm 10^{23}\,cm^{-2}$) pre-shock flow, 
partially covering the X-ray emitting pole, is an additional
complication in interpreting the X-ray spectra and spin modulations of MCVs 
\citep{rosen88,mukai94}. Moreover, the detection of absorption edges of ionised 
Oxygen or Iron in a few IPs \citep{mukai01,demartino08,bernardini12} testifies 
that the accretion flow can also be ionised also in MCVs, besides low mass X-ray binaries.

Thanks to the deep hard X-ray surveys carried out by \INT\ and \Swift\ satellites above 20 keV, 
and the numerous multiwavelength followup \citep[see e.g.][and reference therein]{masetti13}
the number of MCVs increased rapidly in the last years, the majority being identified as magnetics of the IP class, 
with a few polars and a handful of non-magnetic systems \citep{barlow06,bird10,cusumano10,baumgartner13}. The role of these hard X-ray
emitting CVs as dominant contributors to the low-luminosity ($\rm \lesssim 10^{33}\, erg\,s^{-1}$)
X-ray source population in the galactic ridge (GRXE) and Galactic centre 
is a greatly debated topic for a decade 
\citep{muno04,sazonov06,revnivtsev09,revnivtsev11,warwick14,perez15,hailey16}. Whether the low-luminosity component is due to non-magnetic CVs or to a still to be discovered large population of IPs accreting at low rates has yet to be established. This would require higher space densities of MCVs than those, still very
uncertain, obtained from current observations 
\citep[see][and reference therein]{reis13,pretorius14}. 
In this respect, the identification of new systems is important to increase the statistics.  
 
We here present the analysis of simultaneous X-ray and 
optical data of seven recently identified candidates, 
to unambiguously confirm them as magnetics through the
search for X-ray pulsations and the study of their broad-band 
spectral properties. Our source sample consists of 
Swift\,J0706.8+0325, Swift\,J0746.3-1608, PBC\,J0801.2-4625, Swift\,J0927.7-6945, 
Swift\,J0958.0-4208, Swift\,J1701.3-4304, and Swift\,J2113.5+5422 
(hereafter J0706, J0746, J0801, J0927, J0958, J1701, and J2113, respectively). 
Their magnetic candidacy was only based on optical 
spectroscopic follow-ups \citep[][and references therein]
{masetti13,thorstensen13,parisi14,halpern15}. 
 Spectroscopic orbital periods were recently
determined for J0706 \citep[1.7 h;][]{halpern15} and J0746 
\citep[9.4 h;][]{thorstensen13}, suggesting a Polar and a novalike, respectively.
\\

\section{Observations and data reduction} 
\label{sec:obs}

\subsection{\textit{XMM-Newton} observations}

The sources were observed between 2015-06-05 and 2016-04-01 by \XMM\ with 
the European Photo Imaging Cameras \citep[EPIC: PN, MOS1 and MOS2][]{struder01,turner01,denherder01} 
as main instruments, complemented with simultaneous optical monitor \citep[OM,][]{mason01} photometry. 
The details of the observations are reported in Table \ref{tab:observ}. Data were processed using the 
Science Analysis Software (\textsc{SAS}) version 15.0.0 and the latest calibration files available in 2016 February.

\begin{table*}
\caption{Summary of main observation parameters for all instruments. Uncertainties are at $1\sigma$ confidence level.}
\begin{center}
\begin{tabular}{cccccccc}
\hline 
Source Name            & Telescope            & OBSID        & Instrument    & Date        & UT$_{\rm start}$ & T$_{\rm exp}$ $^a$ & Net Source Count Rate\\
Coordinates (J2000)$^{\alpha}$&               &              &        & yyyy-mm-dd      & hh:mm & ks      &    c/s                  \\
\hline
Swift\,J0706.8+0325 & \XMM\                &  0761120501  & EPIC-PN$^b$   & 2015-10-25  & 15:00 & 39.6 & $0.896\pm0.006$ \\
                   &                      &              & EPIC-MOS1$^b$ & 2015-10-25  & 17:03 & 32.0 & $0.237\pm0.003$     \\ 
RA=07:06:48.90	   &                      &              & EPIC-MOS2$^c$ & 2015-10-25  & 14:55 & 39.8 & $0.293\pm0.003$     \\  
Dec=+03:24:47.3    &                      &              & RGS1          & 2015-10-25  & 14:54 & 40.0 & $0.031\pm0.002$      \\
                   &                      &              & RGS2          & 2015-10-25  & 14:54 & 40.0 &  $0.036\pm0.002$    \\                 
                   &                      &              & OM-B$^d$      & 2015-10-25  & 15:01 & 6.6  & 17.52$\pm0.02^e$    \\
                   & \Swift\              &              & BAT$^f$     &             &	      & 7020 & $1.7\pm0.3\times 10^{-5}$ \\ 
\hline              
Swift\,J0746.3-1608& \XMM\                &  0761120401  & EPIC-PN$^c$   & 2016-04-01  & 08:29 & 34.5 & $0.136\pm0.003$ \\
                   &                      &              & EPIC-MOS1$^c$ & 2016-04-01  & 08:23 & 34.7 & $0.035\pm0.001$     \\ 
RA=07:46:17.13     &                      &              & EPIC-MOS2$^c$ & 2016-04-01  & 08:24 & 34.7 & $0.039\pm0.001$\\  
Dec=-16:11:27.8    &                      &              & OM-B$^d$      & 2016-04-01  & 08:30 & 31.0 & 16.33$\pm$0.06$^e$    \\
                   & \Swift\              &              & BAT$^f$     &             &	      & 8710 & $0.105\pm0.03$ \\                         
                   & \Swift\              &              & XRT$^g$     &             &	      &  106 & $0.28\pm0.03$ \\      
\hline
PBC\,J0801.2-4625  & \XMM\                &  0761120301  & EPIC-PN$^h$   & 2015-12-04  & 18:50 & 30.0 & $0.747\pm0.006$ \\
                   &                      &              & EPIC-MOS1$^h$ & 2015-12-04  & 18:27 & 31.7 & $0.192\pm0.003$     \\ 
RA=08:01:17.03     &                      &              & EPIC-MOS2$^h$ & 2015-12-04  & 18:28 & 31.7 & $0.192\pm0.003$     \\  
Dec=-46:23:27.5    &                      &              & OM-V$^i$      & 2015-12-04  & 18:33 & 26.6 & $15.61\pm0.01^e$    \\
                   & \INT\                &              & IBIS/ISGRI$^l$    &             &	      & 3530 & $0.12\pm0.02$ \\                   
\hline  
Swift\,J0927.7-6945 & \XMM\                &  0761120901  & EPIC-PN$^h$   & 2015-06-05  & 01:10 &   31.6 & $0.471\pm0.004$ \\
                   &                      &              & EPIC-MOS1$^h$ & 2015-06-05  & 00:46 &   33.3 & $0.140\pm0.002$     \\ 
RA=09:27:53.12 	   &                      &              & EPIC-MOS2$^h$ & 2015-06-05  & 00:47 &   33.2 & $0.147\pm0.002$     \\  
Dec=-69:44:42.0    &                      &              & OM-B$^d$      & 2015-06-05  & 00:53 &   26.4 & 16.10$\pm$0.02$^e$    \\
                   & \Swift\              &              & BAT$^f$     &             &	      & 11700  & $2.0\pm0.2\times 10^{-5}$ \\ 
\hline              
Swift\,J0958.0-4208 & \XMM\                &  0761120101  & EPIC-PN$^h$   & 2015-05-14  & 04:56 & 37.9 & $0.684\pm0.007$ \\
                   &                      &              & EPIC-MOS1$^h$ & 2015-05-14  & 04:33 & 38.2 & $0.203\pm0.003$     \\ 
RA=09:57:50.64     &                      &              & EPIC-MOS2$^h$ & 2015-05-14  & 04:34 & 38.1 & $0.204\pm0.004$     \\  
Dec=-42:08:35.5    &                      &              & OM-B$^d$      & 2015-05-14  & 04:39 & 34.0 & $15.61\pm0.01^e$    \\
                   & \Swift\              &              & BAT$^f$     &             &	      & 9550 & $1.7\pm0.3\times 10^{-5}$ \\ 
\hline              
Swift\,J1701.3-4304 & \XMM\                &  0761120701  & EPIC-PN$^h$   & 2015-09-22  & 15:25 & 30.0 & $0.896\pm0.006$ \\
                   &                      &              & EPIC-MOS1$^h$ & 2015-09-22  & 15:02 & 31.7 & $0.237\pm0.003$     \\ 
RA=17:01:28.15     &                      &              & EPIC-MOS2$^h$ & 2015-09-22  & 15:02 & 31.7 & $0.293\pm0.003$     \\  
Dec=-43:06:12.3    &                      &              & OM-B$^d$      & 2015-09-22  & 15:08 & 26.6 & $16.98\pm0.01^e$    \\
                   & \Swift\              &              & BAT$^f$     &             &	      & 8050  & $2.2\pm0.4\times 10^{-5}$ \\ 
\hline              

Swift\,J2113.5+5422 & \XMM\                &    			& EPIC-PN$^h$   & 2015-11-17  & 17:28 & 36.3 & $1.021\pm0.006$ \\
                   &                      &              & EPIC-MOS1$^h$ & 2015-11-17  & 17:05 & 38.0 & $0.323\pm0.003$     \\ 
RA=21:13:35.40     &                      &              & EPIC-MOS2$^h$ & 2015-11-17  & 17:05 & 38.0 & $0.342\pm0.003$     \\  
Dec=+54:22:32.8    &                      &              & OM-V$^i$      & 2015-11-17  & 17:04 & 31.0 & $18.65\pm0.07^e$    \\
                   & \Swift\              &              & BAT$^f$       &             &	      & 13000 & $1.6\pm0.2\times 10^{-5}$ \\ 
\hline              
\end{tabular}
\label{tab:observ}
\end{center}
\begin{flushleft}
$^{\alpha}$ Coordinates of the optical counterpart. \\
$^a$ Net exposure time.\\
$^b$ Large window mode (thin filter applied). \\
$^c$ Small window mode (thin filter applied). \\
$^d$ Fast window mode. The central wavelength of the B filter is 4500 \AA. \\ 
$^e$ OM instrumental magnitude.\\
$^f$ All available pointings collected from 2004 December to 2010 September are summed together. \\
$^g$ All observations available between 2009-08-28 and 2014-06-20 (observation id: 49197, 38960, 40698, 41163, and 90159) are summed together. \\
$^h$ Full frame mode (thin filter applied). \\
$^i$ Fast window mode. The central wavelength of the V filter is 5430 \AA. \\
$^l$ All available pointings collected from 2003-04-05 to 2014-12-20 are summed together. \\
\end{flushleft}
\end{table*}

\subsubsection{The EPIC and RGS data}

For the EPIC data, source photon event lists and spectra were extracted 
from a circular region of radius 40 arcsec. 
Background photons were taken from a nearby region of the sky 
clear from sources contamination, in the same CCD where the source lies, avoiding CCD gaps.
For the spectral analysis, high particle background epochs were 
removed in all instruments, while 
for the timing analysis, we used the whole data set when possible.

Background-subtracted light curves were also generated with the task \textsc{epiclccorr} in 
the ranges $0.3-12$ keV, $0.3-1$ keV, $1-3$ keV, $3-5$ keV and $5-12$ keV, 
with different bin size, depending on the source and background flux. 
The event arrival times were corrected to Solar system barycenter by using the 
task \textsc{barycen}. 
Spectra were rebinned before fitting with the task \textsc{specgroup}, in order 
to have a minimum of 50 and 25 counts in each bin for PN and MOSs, respectively, 
and to prevent oversampling of the energy resolution by more than a factor of 
three. 
Spectra were also extracted as a function of the source rotational phase 
and of the orbital phase, when applicable. 
The response matrix and the ancillary files were generated using the tasks 
\textsc{rmfgen} and \textsc{arfgen}. 
The spectral fits were made with \textsc{Xspec} version 12.8.2 \citep{arnaud96}.

All sources except J0706, are too faint for a 
meaningful analysis with the RGS. For that source, the RGS1 and RGS2 spectra 
from the standard \textsc{SAS} pipeline were used (Table \ref{tab:observ}).

\subsubsection{The Optical Monitor photometry}

In all observations the OM was operated in fast window mode using
the B-band (3900--4900 \AA) filter, except for 
J0801 and J2113 where the V-band (5100--5800 \AA) filter was used due to bright
source in their field of view. 
The background subtracted light curves were generated with the 
task \textsc{omfchain}, with different bin times ($\sim1-10$ s) depending 
on the source flux, and then barycentric corrections were applied. 

\subsection{The \Swift\ observations}

BAT has built up an all-sky map of the hard X-ray sky (14--195 keV), thanks to its wide field of view.
For all sources, with the exception of J0801 for which 
no spectrum is available, 
we used the archival eight-channel spectra from the first 70 months of 
monitoring \citep{baumgartner13} available at the NASA GSFC 
site\footnote{http://swift.gsfc.nasa.gov/results/bs70mon/}.  The sources
under study are all detected up to energies $\sim80$ keV up to which
we restrict our spectral analysis.

We also used XRT archival data to study the long term variability of two sources (J0746 and J0801). For this scope we used the lightcurves and spectra from the XRT products generator \citep{Evans09} at the UK \Swift\ Science Data Centre at the University of Leicester\footnote{http://www.swift.ac.uk/}

\subsection{The \INT\ observations}

For J0801 we extracted the \INT/IBIS spectrum from all the 
1845 public \INT\ pointing with RA$\in [109,135]$ and Dec 
$\in [-53,36]$. The Offline Analysis Software v. 10.2 distributed
by the ISDC \citep{isdc} was used.
We built a response with four logarithmically spaced bins 
between 20 and 100 keV and extracted the source flux using the standard 
spectral extraction. We derived a sky model from a mosaicked significance 
map in the 20-100 keV band, which includes 39 sources, and an average mission background, 
retrieved from the calibration files. The spectral analysis of J0801 is further restricted below 80 keV.

\subsection{Optical ground based photometry}
\label{opticalground}

The detection of deep X-ray and optical eclipses in 
J1701 (Section \ref{J1701}) motivated the search for its 
orbital period using optical ground-based photometry.

J1701 was repeatedly observed during ten days with the 1-m network of robotic optical telescopes in the Las Cumbres Observatory (LCO), from 2016-06-06 20:29:08 UT to 2016-06-16 03:34:33 UT. Six telescopes were used, located at three sites, Cerro Tololo (Chile), South African Astronomical Observatory (SAAO), Sutherland (South Africa) and Siding Spring (Australia). Imaging was obtained in the Sloan Digital Sky Survey (SDSS) $g^{\prime}$-band with exposure times of 40 s. The coverage was typically 40 minutes about three times in 24 hours. A total of 328 usable images were acquired. The automatic Las Cumbres Observatory (LCO) pipeline \emph{BANZAI} provides users with bias and flat-field corrected science images. Additional photometry was acquired during seven nights for about 5 hours consecutively 
from 2016-07-27 00:05:13\,UT to 2016-08-07 06:08:00\,UT 
with the robotic 0.6 m INAF {\it REM} telescope 
in La Silla, Chile \citep{Zerbi04}. 
The telescope is equipped with the ROSS2 camera\footnote{http://www.rem.inaf.it}
that performs simultaneous exposures in the Sloan filters g', r', i' and z' and with
the REMIR camera \citep{Conconi04} covering simultaneously the near-IR 
band. Integration times were 150\,s for all optical filters and a
dithering of 5 exposures of 60\,s each was used for the REMIR J-band.
Both ROSS2 and REMIR data sets were reduced using standard routines of
IRAF to perform  bias and flat-field  corrections. Due to the low response
of the z' filter the corresponding images have not been analysed.
For the REMIR observations  the five dithered images were merged
into a single frame.
A total of 483 useful science images were secured in each g',r',i'  filters 
while 251 were obtained in the J band. 

 Aperture photometry was performed on both LCO and {\it REM} data sets
optimizing aperture radius and sky subtraction by adopting annuli of different
sizes.  Comparison stars  were used to check  and to correct for variable
sky conditions.  {\it REM}/ROSS2 and REMIR photometry was calibrated
using the Sloan standards SA\,104\,428, SA\,114\,531 and SA\,093\,317 
observed each night in July and August, respectively, whose near-IR magnitudes 
are also tabulated in the 2MASS catalogue \citep{skrutskie06}
\footnote{http://www.ipac.caltech.edu/2mass/}.
The LCO photometry was instead calibrated using three comparison stars with known g' magnitudes. Both data sets in the g' band agree within 9 per cent.
Average magnitudes are: g'=$16.50\pm0.02$, r'=$15.389\pm0.009$, i'=$14.999\pm0.008$, and J=$13.31\pm0.02$ mag.

J1701 is also identified in the {\it AAVSO Photometric All-Sky 
Survey (APASS)}\footnote{http://www.aavso.org/apass } as  
GDS\,J1701281-430612 \citep{henden16}. 
It was recently observed by AAVSO from 2016-09-20 23:15:19\,UT
to 2016-10-27 10:46:14\,UT 
with several filters. For our purposes, we selected the most dense coverage in the V-band, including clear (unfiltered) reduced to V-band observations, 
totaling 3943 photometric measures. The average V magnitude is $16.05\pm0.04$. 

\noindent All light curves were corrected to the Solar system barycentre.

\section{Data analysis}
\label{sec:danalysis}

In this section, we describe the common procedures adopted for all sources 
to perform the timing and spectral analysis. We present the results, for 
each source individually, in Section \ref{sec:Results}.

\subsection{Timing analysis} 

Light curves in the 0.3--12 keV range (PN and MOS) and in the 
optical OM B or V bands
were first inspected for short and long-term variability 
(Figure \ref{fig:orbital}). Then power spectra were computed 
to identify periodic signals in the two energy domains. 
A phase-fitting technique \citep[see e.g.][for details on the technique]{dallosso03}, was then used for an accurate period determination of the stronger signal. In case of additional significant peaks, 
the periods were determined using the \textsc{FTOOLS}\footnote{ 
http://heasarc.gsfc.nasa.gov/ftools/}
task \textsc{efsearch} \citep[][]{blackburn95}.

We interpreted the pulsation associated with the main peak in the 
power spectrum as the WD spin period (P$^{X}_{\omega}$), the lower signals
to sidebands (P$^{X}_{side}$) and, when present, to the orbital
(P$^{X,lc}_{\Omega}$) periods. The most common and 
stronger sideband observed in IPs is the
beat ($\omega - \Omega$) between the spin and the
orbital  ($\Omega$) periods and, when present, we used it 
to estimate the orbital period.

For those sources displaying detectable orbital variability, a fit 
using one or more sinusoids (the fundamental plus possible harmonics) was performed to estimate the orbital period in the X-ray (P$^{X,lc}_{\Omega}$) and
optical (P$^{opt}_{\Omega}$) bands. The two procedures were verified to
give consistent results.

Spectral variations along the spin period  
were first inspected by folding at P$^{X}_{\omega}$ 
the background subtracted source light curves extracted 
in five energy bands: 0.3--1, 1--3, 3--5, 5--12 keV and 
computing the hardness ratios (HR) defined as the count rate ratio 
in each phase bins between two selected energy ranges. A quantitative
analysis (Table \ref{tab:time}, second column) was then 
performed by fitting the pulses in the above bands with a Fourier sine series truncated at the highest significant
harmonic. For each energy interval, the pulsed fraction was computed as:
$\rm PF=(A_{max}-A_{min})/(A_{max}+A_{min})$, where $\rm A_{max}$ and $\rm
A_{min}$ are respectively the maximum and minimum amplitudes of the sinusoid
at the fundamental frequency. 

\begin{table*}
\caption{Timing properties of the source sample. Uncertainties are at $1\sigma$ confidence level.  
From left to right: P$^X_{\omega}$ (X-ray spin period);
 P$^X_{side}$ (X-ray sidebands); P$^{X,side}_{\Omega}$ (orbital period derived from 
X-ray sidebands); 
P$^{X,lc}_{\Omega}$ (orbital period derived from X-ray light curve fitting); 
A$^X_{\omega}$/A$^X_{side}$ (spin to sideband X-ray amplitude ratio);
P$^{opt}_{\omega}$ (optical spin period from OM); P$^{opt}_{\Omega}$ 
(optical orbital period as derived from OM or ground-based observations); 
P$^{A}_{\Omega}$(adopted orbital period in this work with its reference 
in parenthesis). 
}
{\small
\begin{center}
\tabcolsep=0.05cm
\begin{tabular}{ccccccccc}
\hline 
   &                   &         & &               & &  &  \\
Source  & P$^X_{\omega}$ &  P$^X_{side}$  & P$^{X,side}_{\Omega}$ & P$^{X,lc}_{\Omega}$ 
& A$^X_{\omega}$/A$^X_{Side}$ & P$^{opt}_{\omega}$ &  P$^{opt}_{\Omega}$&  \textbf{P$^{A}_{\Omega}$}\\
        & s & s  & h & h &   & s   & h & h  \\
\hline 
\\
J0706 & $6135\pm5$ & -  & -   & P$_{\omega}=$P$_{\Omega}$ & - & -  
& -  & 1.7018$\pm$0.0003 [1] \\

J0746 & $^a$  & - & -   & -  &  - & -  &  5.03$\pm$0.10 $^{b}$  &  9.3841$\pm$0.0002 [2] \\

J0801 & $1310.9\pm1.5$  & - & - & - &  - & $1306.3\pm0.9$ & - & - \\

J0927  & $1033.54\pm0.51$ & $1093.4\pm6.5$ ($\omega-\Omega$) & $5.25\pm0.45$  & $5.15\pm0.10$ & $\sim2.5$ & $1030.6\pm0.9$ & - &  
5.15$\pm$0.10 [3] \\

J0958 & $296.22\pm0.05$ & - & - & - & - &  - & - & -  \\

J1701 & 1859$\pm$3  & -  & - & -  &  -  & $1858.67\pm0.02^c$  
& 12.81662$\pm$0.00012$^d$  
& 12.81662$\pm$0.00012 [3]  \\
J2113 & 1265.6$\pm$4.5  & 1373.8$\pm$2.6 ($\omega-\Omega$) &  $4.46\pm0.10$  
& 4.02$\pm$0.10  & $\sim1.1$ &  - & 3.63$\pm$0.11  & 4.02$\pm$0.10 [3]   \\
\hline 
\end{tabular}
\label{tab:time}
\end{center}}
\begin{flushleft}
$^a$ Short-term variability with timescale of about 2700 s is present in the first half of the observation only. \\
$^b$ Period derived from OM data (=2P$_{\Omega}$). \\
$^c$ Spin period derived from AAVSO V-band photometry. From OM we get $1857\pm6$ s. \\
$^d$ Orbital period derived from multi-site optical photometry. \\
\noindent [1] Halpern \& Thorstensen 2015; [2] Thorstensen \& Halpern 2013; [3] This work. \\
\end{flushleft}
\end{table*}

\subsection{Spectral analysis}

The study of the broad-band (0.3--80 keV) spectral properties of each 
source was first accomplished on the grand-average PN, MOS1 and MOS2 
spectra together with the corresponding 
\Swift-BAT or \INT-IBIS spectra fitted simultaneously. 
Calibration discrepancies between 
different instruments were taken into account by including in the fits
an inter-calibration constant which only for the PN is fixed to one. 
This constant also accounts for the non simultaneity of the BAT(IBIS) 
spectra with those from the EPIC cameras and so for possible source long-term variability. All model parameters were linked between different instruments expect the multiplicative constants.

The sources under study display thermal spectra and most of them also
show emission features at the iron complex. These are generally observed in the MCVs which have multi-temperature spectra locally
absorbed by high dense cold material 
\citep[see e.g.][]{Done95,EzukaIshida99,beardmore00,
demartino04,anzolin09,bernardini12,mukai15}. 
Therefore, the spectra were fitted with a composite model consisting of 
one or more optically thin plasma components (\textsc{bremss}, \textsc{mekal} 
or \textsc{cemekl} in \textsc{xspec}), with metal abundances (A$_{\rm Z}$) with respect to Solar from \cite{wilms00} left free to vary,  
absorbed by a total (\textsc{phabs}) and one or two partial (\textsc{pcfabs}) covering columns and a narrow Gaussian line 
fixed at 6.4 keV to account for the fluorescent Fe K$_{\alpha}$ feature. 
When needed, a blackbody component (\textsc{bbody}) was 
also included to model a soft excess in the residuals. 
Despite the presence of the 6.4 keV feature the inclusion of a reflection component (\textsc{reflect}) does not improve the fit quality in any of the sources under study, indicating that higher quality
spectra above 10 keV are needed \citep[see also][]{mukai15}.
Additionally, to obtain a more reliable
estimate of the mass of the accreting WD, the broad-band continuum spectra 
were also fitted with the more 
physical post-shock region (PSR) model developed by \cite{suleimanov05}.
This model  takes into account  both temperature and gravity gradients 
within the PSR in bremsstrahlung dominated regimes where cyclotron cooling is 
negligible, and thus applicable to those sources identified as IPs and, with
caution, also to a low-field polar (see Sect.\,4.1.2).
 To this purpose, for those
sources showing soft complexities, the spectra were 
analysed above 1--3 keV to avoid the effect of absorption and 
heavily rebinned to smear emission lines, 
except the iron line complex, which is accounted for by a broad Gaussian.

To unveil the role of spectral parameters in producing the 
X-ray spin(orbital) modulations, a pulse-phase spectroscopy (PPS) analysis 
was performed on the EPIC spectra extracted at spin(orbital) maximum and minimum,
which were fitted separately. For each source we adopted the spectral 
models and parameters obtained from the phase average spectral fits, 
fixing the column density of the total 
absorber and the metal abundance at their best fit values. 
All other parameters were left free to vary in the spin PPS analysis. 
In the orbital PPS analysis only the partial covering absorbers and the 
\textsc{mekal(s)} normalization(s) ($\rm n_c$ and $\rm n_h$) were left free to vary when this extra free component is statistically significant (F-test verified).

\section{Results}
\label{sec:Results}

\subsection{Swift\,J0706.8+0325}
\label{J0706}

J0706 has been recently identified by \cite{halpern15}, where it appears as PBC J0706.7+0327, as a short (1.7 h) orbital period CV 
and proposed as a polar based on its emission line variations and large amplitude optical modulation.

\subsubsection{Timing analysis}

\begin{figure*}
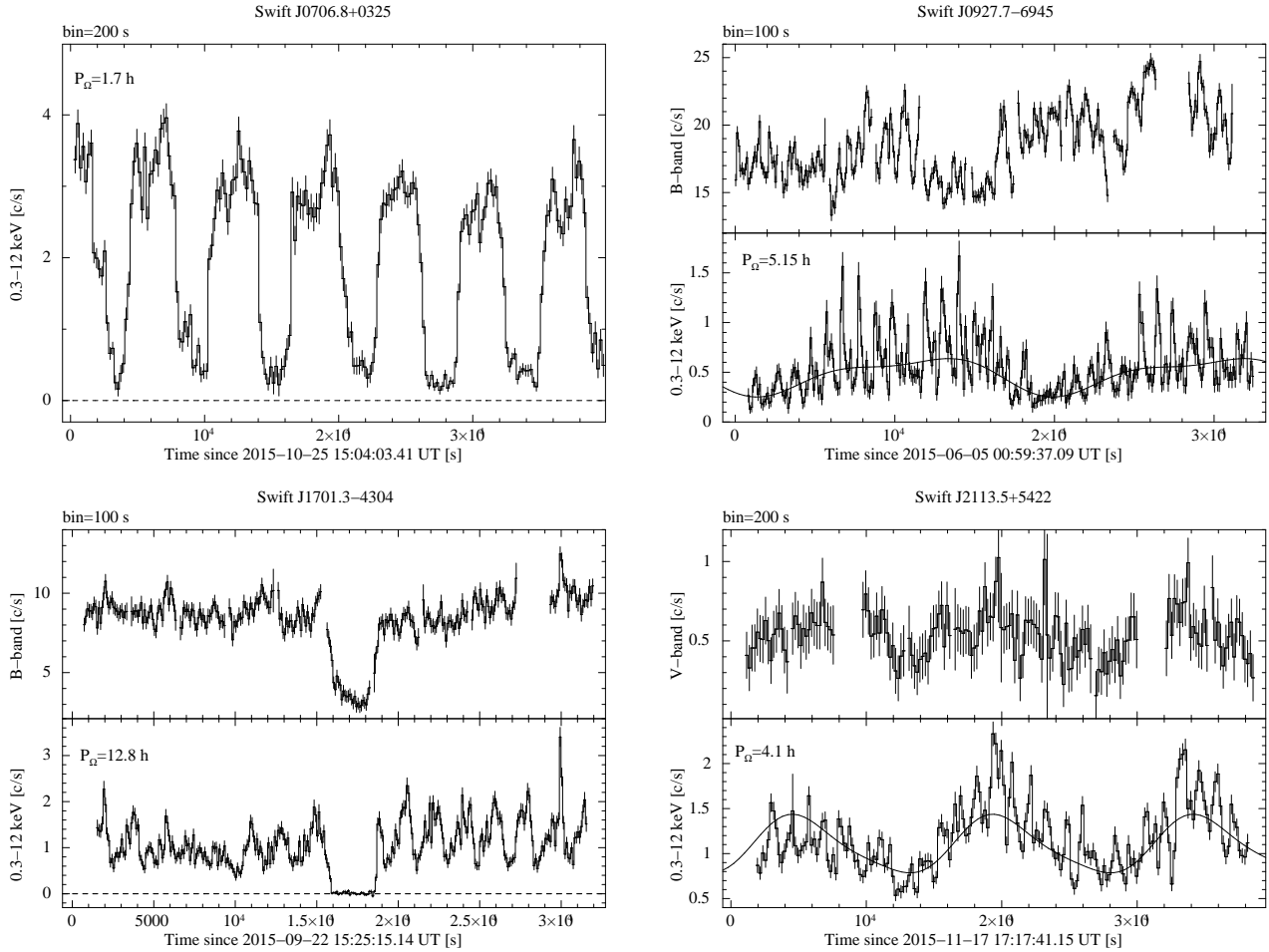

\begin{center}
\begin{tabular}{cc}
\includegraphics[angle=270,width=3.3in]{j0706_orbital_lc.eps} & \includegraphics[angle=270,width=3.3in]{j0927_orbital_lc.eps}\\ 
\includegraphics[angle=270,width=3.3in]{j1701_orbital_lc.eps} & \includegraphics[angle=270,width=3.3in]{j2113_orbital_lc.eps}\\ 
\end{tabular}
\caption{Optical OM (upper panels) and X-ray PN (lower panels)
light curves for those sources showing orbital modulation (or long-term features). 
\textit{Top left}: The X-ray light curve of the polar 
J0706 is modulated at ($P_{\Omega}=P_{\omega}=1.7$ h). The dashed line marks the zero counts level.
No useful OM data are available for this source. 
\textit{Top Right}: The X-ray light curve of J0927 shows 
modulation at both spin and orbital periods. The latter is depicted for plotting purposes  
with a solid line and is a model consisting of two sinusoids (the 5.15 h and its first 
harmonic). 
\textit{Bottom Left}: The X-ray light curve 
of J1701 shows a total eclipse, while in the optical band the 
eclipse is partial. The modulation at the spin period is clear in both bands. 
The dashed line marks the zero counts level. \textit{Bottom Right}: The X-ray 
light curve of J2113 showing short-term variability as well as
an orbital modulation depicted by a solid line representing a sinusoid at the
period of 4.02 h. The optical V band is also modulated at the
orbital period.} 
\label{fig:orbital}
\end{center}
\end{figure*}

The X-ray light curve covers about seven cycles of an almost on-off modulation, 
although the count rate does not drop to zero at minima 
(Figure \ref{fig:orbital}). From a multi-sinusoidal fit to reproduce the shape of the modulation, we get a period of $6135\pm5$ s, which is within $2\sigma$ consistent with the more accurate orbital period 
($6126.36\pm0.10$ s) determined by \cite{halpern15} with optical spectroscopy (Table \ref{tab:time}).
This allows to interpret the X-ray period as the spin of the WD which is  
synchronised with the binary orbit. Thus, J0706 is unambiguously identified
as a polar system. 
The OM B-band photometry being affected by 
technical problems is not used in the present study.

\begin{figure*}
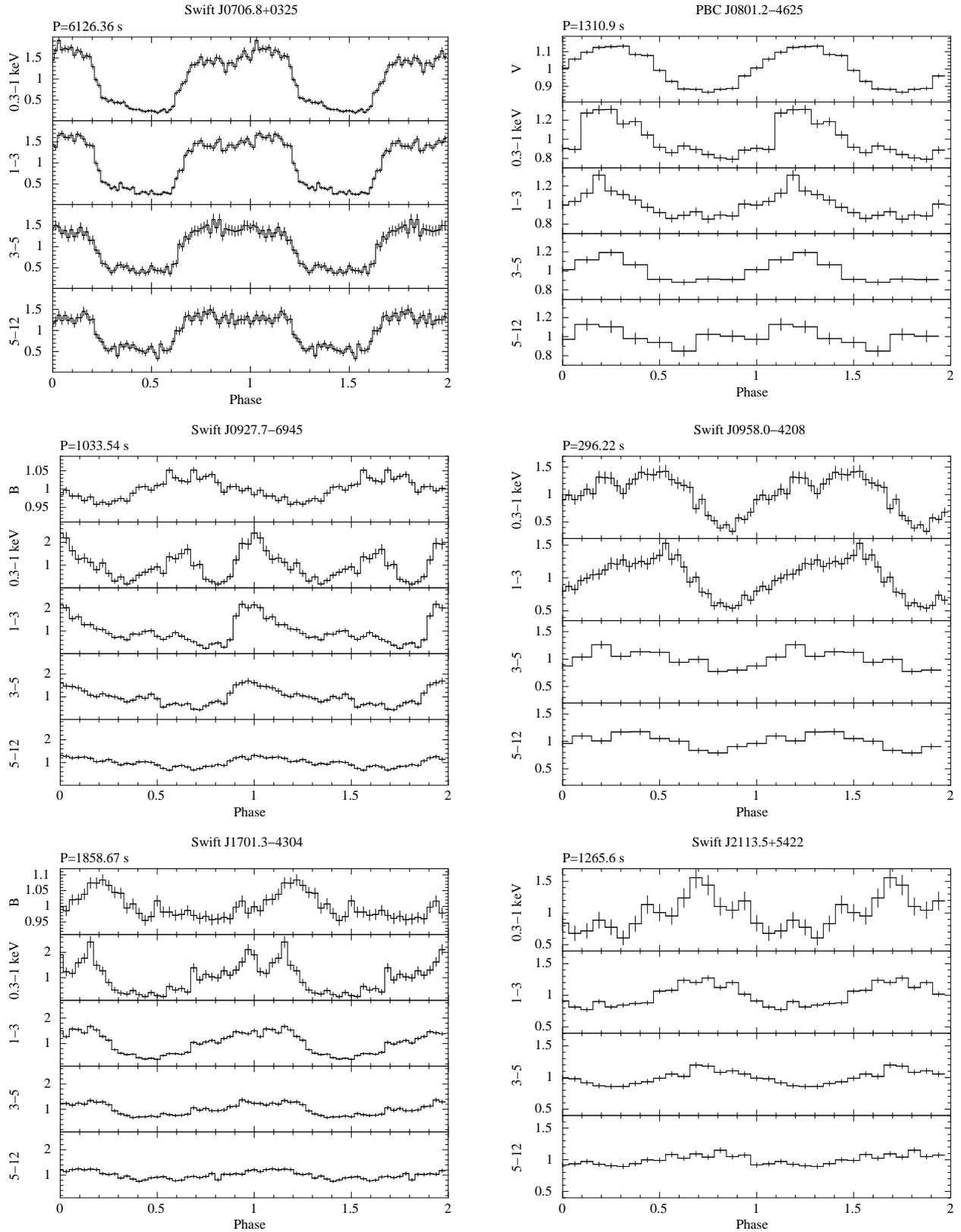

\begin{center}
\begin{tabular}{cc}
\includegraphics[angle=270,width=3.3in]{j07068_pfall.eps} & \includegraphics[angle=270,width=3.3in]{j0801_pfall.eps}\\ 
\includegraphics[angle=270,width=3.3in]{j09278_pfall.eps} & \includegraphics[angle=270,width=3.3in]{j0958_pfall.eps}\\
\includegraphics[angle=270,width=3.3in]{J1701_pfall.eps} & \includegraphics[angle=270,width=3.3in]{j2113_pfall.eps}\\  
\end{tabular}
\caption{
X-ray (PN) pulse profiles at different energy intervals. Energy increases \textit{from top to bottom}. Two pulse cycles are shown for plotting purposes. 
The folding period is also reported on the top left of each figure. The reference folding time for each source is the integer of the observation starting time (Table \ref{tab:observ}) with the exception of J0706, where is MJD=56681.8996 \citep{halpern15}.
For all sources, the X-ray PF decreases as the energy increases (Table \ref{tab:pf}). Optical pulse profiles are also shown when the source displays optical pulses. Since the optical pulse is less intense than the X-ray pulse, it is plotted on a different scale.}
\label{fig:pulseprof}
\end{center}
\end{figure*}

The X-ray light curves at different energies  were folded using the spectroscopic ephemeris of \cite{halpern15}.The modulation shows a broad asymmetric maximum extending $\sim$0.6 in phase. It is structured below 3 keV with a lower count rate during the first half of the maximum and a dip at $\phi\sim0.95$ followed by a higher rate lasting $\sim0.2$ in phase. It is instead  
flatter for E$>3$ keV (Figure \ref{fig:pulseprof}). 
The faint phase (pulse minimum) is slightly flatter above 3 keV, while it shows 
a smooth decline below it. The PF
varies from 75 per cent in the 0.3-1 keV band to 45 per cent in the 
5--12 keV band (Table \ref{tab:pf}). 
The modulation reveals the typical bright
and faint phases seen in the polars, which are produced by the
accretion flow above the main (or upper) pole which comes into (bright
phase) and out of (faint phase) view if the magnetic and rotation axes 
are offset by an angle $\beta$ defined as the magnetic colatitude, and
shifted in azimuth ($\psi$) \citep[see][]{Cropper88}. 
The non-zero count rate during the faint phase
could suggest either that the accreting upper pole does not completely
disappear behind the WD limb or that a second emitting region is present. 
The length of the 
faint phase can be used to restrict the range of values of the 
binary inclination $i$ and colatitude $\beta$ of the main pole 
\citep{Cropper90}. The lack of eclipses implies $\leq 75^o$.
These give $41^o \lesssim \beta \lesssim 87^o$ for $10^o \lesssim i \lesssim 75^o$.
The energy dependence of the dip and pre-dip maximum 
is reminiscent of absorption from the accretion stream 
at the threading of magnetic lines as seen in other polars 
\citep[e.g.][]{ramsay07,ramsay09,bernardini14}, implying 
$i > \beta$. This further restricts the binary inclination
to $ 41^o \lesssim i \lesssim 75^o$. Additionally, the fast rise/decay 
of the bright phase can be used to estimate the extent of the emitting region.
However, both bright and faint phases suffer from cycle-to-cycle variability, 
thus a rise time of $\sim600$ s is estimated from the more symmetric profiles 
observed at the $\rm 3^{rd}$,$\rm 4^{th}$ and $\rm 5^{th}$ cycles. A similar decay time is observed in these cycles.
This gives an azimuthal extent of the spot of 35$^o$, assuming no lateral extent. 

\begin{table}
\begin{center}
\tabcolsep=0.1cm
\caption{Pulsed fraction vs energy. Results refer to the fundamental frequency ($\omega$, Table \ref{tab:time}, column 2), n is the number of sinusoids used to fit the data. Uncertainties are at $1\sigma$ confidence level.}
\begin{tabular}{cccccccc}
\hline     
Source        & \multicolumn{5}{c}{Pulsed Fraction}& n \\   
  &0.3--1 keV& 1--3 keV  & 3--5 keV  & 5--12 keV   & 0.3--12 keV & \\
        & \%   &  \%  & \%   & \%      &  \%  & \\
\hline 
\\
J0706$^a$  & 75(1) & 71(1)     & 60(2)     & 45(2)    & 68(1)    & 3  \\              
J0927$^b$  & 37(4) & 52(3)     & 40(2)     & 23(2)    & 33(1)    & 3  \\
J0958$^c$  & 38(2) & 30(2)     & 23(4)     & 7(2)     & 20(1)    & 2  \\
J0801$^d$  & 20(2) & 15(3)     & 12(6)     & 9(6)     & 15(2)    & 1  \\
J1701      & 70(3) & 55(2)     & 28(2)     & 18(2)    & 36(1)    & 2  \\
J2113      & 31(5) & 22(1)     & 14(1)     & 9(1)     & 15(1)    & 1  \\
\hline 
\end{tabular}
\label{tab:pf}
\end{center}
$^a$ The PF is calculated using the more accurate optical spectroscopy period (6126.36 s) from Halpern \& Thorstensen 2015.\\
$^b$ For E$=0.3-1$ keV only, the first harmonic has the highest PF ($69\pm4$ per cent).\\
$^c$ Results refer to the first 26 ks of the pointing only. The last 12 ks are removed because of high background contamination.\\
$^d$ Results refer to the second half of the pointing only, where the background level is low. \\
\end{table}

\subsubsection{Spectral analysis}

\begin{figure*}
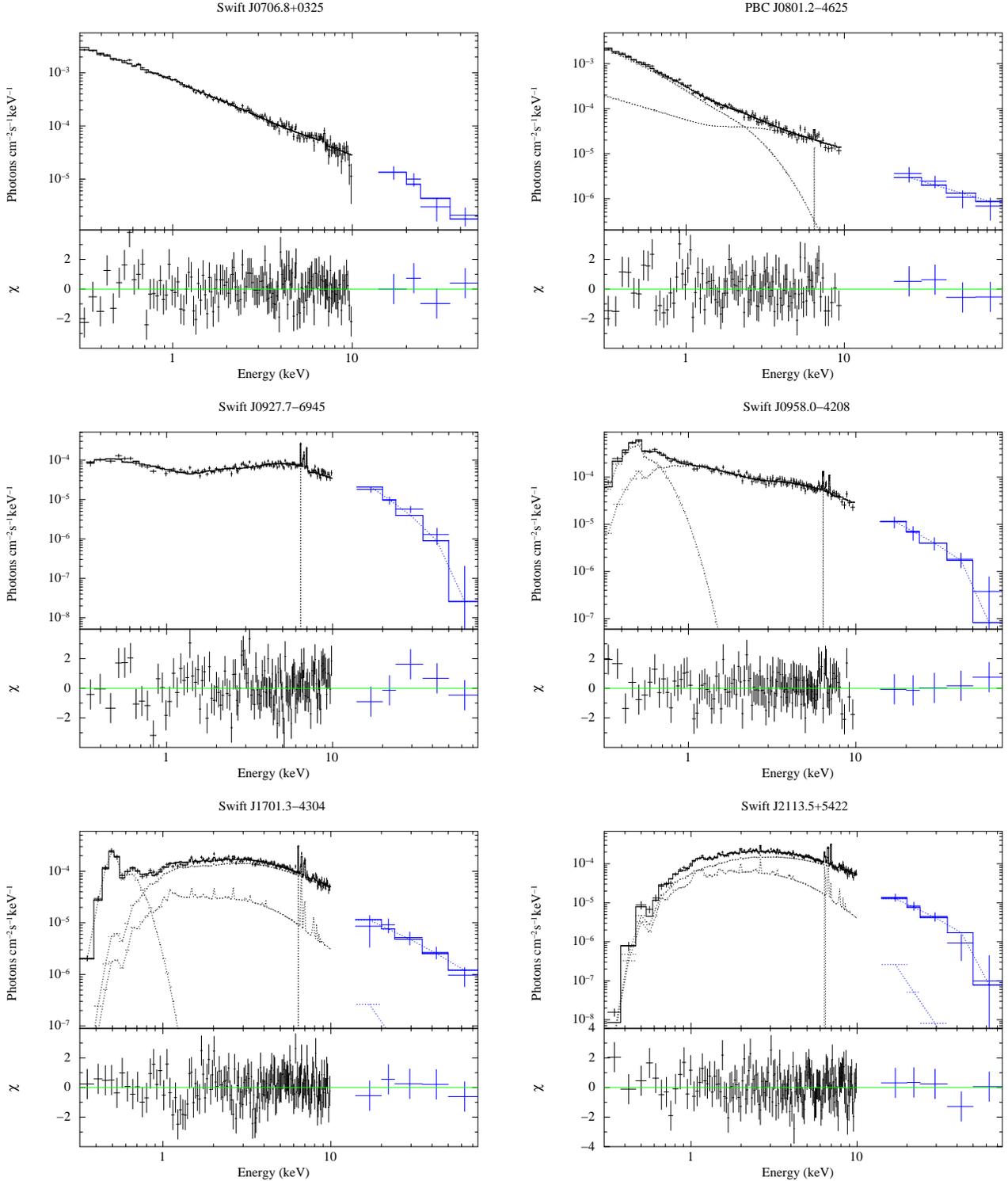

\begin{center}
\begin{tabular}{cc}
\includegraphics[angle=270,width=3.3in]{j0706ufs.eps} & \includegraphics[angle=270,width=3.3in]{j0801ufs.eps}\\ 
\includegraphics[angle=270,width=3.3in]{j0927ufs.eps} & \includegraphics[angle=270,width=3.3in]{j0958ufs.eps}\\
\includegraphics[angle=270,width=3.3in]{j1701ufs.eps} & \includegraphics[angle=270,width=3.3in]{j2113ufs.eps}\\  
\end{tabular}
\caption{Broadband unfolded spectra. Post fit residuals are shown in the \textit{lower panels}. Black points are \XMM/PN data (0.3--10 keV), blue points are \Swift/BAT data (15--80 keV) or \INT/IBIS data (20--100 keV) for J0801 only. The fit is made simultaneously on all EPIC cameras, but for sake of readability, we only show PN data. The dotted lines mark individual model components, while the solid line is the composite model.}
\label{fig:avrspec}
\end{center}
\end{figure*}

While a simple optically thin plasma with kT$=11.5\pm0.7$ keV and $\rm A_Z=0.4\pm0.1$ absorbed by only a partial covering column 
well fit the average broad-band spectrum ($\chi^2=1.06$, 336 d.o.f.), a
multi-temperature plasma \textsc{cemekl} with $\rm \alpha=1.0$ (fixed),
$\rm kT_{max}=46\pm4$ keV, and with the same
underabundance gives the same $\chi^2$ and alleviates a problem of the
inter-calibration  constant between the EPIC and BAT spectra. Despite this, 
the partial covering parameters are less constrained (Table \ref{tab:averagespec} and Figure \ref{fig:avrspec}).
The source spectrum does not require either a soft blackbody component or
a Gaussian for the 6.4 keV line. Furthermore, only a $3\sigma$ upper 
limit to the hydrogen  column density of the total absorber  
is derived, a factor $\sim30$ lower than the total value in the 
source direction \citep{kalberla05}.
This suggests that the source distance is small as
also indicated by its large proper motion \citep{skinner14,halpern15}.
To independently verify that the post-shock plasma is not isothermal 
we performed a joint EPIC+RGS fit using the one-T model with the temperature, abundance, and absorbing partial covering column 
parameters fixed to the values obtained in the EPIC+\INT/IBIS fit. 
The new fit clearly shows an excess at $\sim$0.57 keV,
and probably also around 0.65 keV.  
Allowing for an additional simple absorber and narrow Gaussians, we measure 
a line at 0.573 keV
($=21.647$ \AA) with an equivalent width EW$=15\pm5$ eV at 90 per cent confidence, and 
less robustly a probable line at 0.65 keV. These are identified 
as the He-like triplet of \textsc{OVII} and H-like line of \textsc{OVIII},  
neither  of which is in the best-fit 1-T model.  These lines
indicate the presence of plasma at $\sim0.4-0.6$ keV, likely forming at
the base of the post-shock region. We conclude that the shock plasma is indeed not isothermal and consider the \textsc{cemekl} model the best description 
of the data. 
Although the PSR model is not applicable to polars, J0706 is likely a low-field Polar 
(Sect.\,4.1.3) and thus cyclotron cooling should not strongly affect the 
PSR structure. With this note of caution, 
we consider the derived WD
mass from PSR model a lower limit which results: 
$\rm M_{WD}\geq0.79\,M_{\odot}$  (Table \ref{tab:wdmass}).

Both parameters of the multi-temperature plasma decrease at orbital minimum 
as expected in a polar system, where the main accretion spot comes in and out of view along the orbital cycle. The absorption component is poorly constrained and we are unable to assess whether this contribution is important at the orbital minimum (Table \ref{tab:pps_all}).  

\begin{table*}
\caption{Model parameters of the best fit models to the averaged broad-band spectra
of the sources under study. The absorbed 0.3--10 keV and unabsorbed bolometric 
(0.01--200 keV) fluxes are reported in the last two columns. 
Uncertainties are at $1\sigma$ confidence level.}
{\small
\begin{center}
\tabcolsep=0.03cm
\begin{tabular}{ccccccccccccccccc}
\hline 
\\
source   & mod.     & N$_{\rm H_{Ph}}$  & N$_{\rm H_{Pc1}}$  &  cvf & N$_{\rm H_{Pc2}}$ &  cvf  & kT$_{\rm BB}$ & kT$_{\rm c}$     & kT$_{\rm h}$       & n$_{\rm BB}^{\alpha}$ & n$_{\rm c}$     & n$_{\rm h}$      & A$_{\rm Z}$  & EW  &  F$_{0.3-10}$ &   F$_{\rm X,bol}$    \\

        &          &   $10^{22}$  & $10^{22}$ &      &    $10^{22}$  &    &           &              &                 & $10^{-3}$   & $10^{-3}$       &  $10^{-3}$       &      &    & $10^{-12}$  &    $10^{-11}$      \\

& $\chi^2$/dof   & cm$^{-2}$ & cm$^{-2}$ & \% & cm$^{-2}$ & \%  & eV &  keV & keV & & & & & keV & erg/cm$^2$/s & erg/cm$^2$/s \\         
\hline 
\\
J0706 & cemek$^a$ & $\leq0.01^b$ & - & - & $87\pm_{34}^{55}$ & $33\pm_{10}^{15}$ & - & -  & $46\pm4$ &  -  & - & $11.0\pm_{1.5}^{3.4}$  & $0.4(1)$  & $\leq0.20^b$ & $6.6(3)$ & $\sim1.9$ \\
&1.05/336&&&&&&&&&&&&&&&\\                                                                                    
J0746$^c$ &mek & $\leq0.03^b$ & - & - & - & - & - & $7.4(7)$  & - &  -  & $0.21(1)$ & -  & $2.1(5)$  & - & $0.46(2)$ & $\sim0.06$  \\
&1.10/125&&&&&&&&&&&&&&&\\
J0746$^d$ &A:mek & $0.06(2)$ & - & - & - & - & - & -  & $39\pm_{11}^{25}$ &  -  & - & $4.6(5)$  & 1 (fix)  & - & $8.2(3)$ & $\sim2.1$  \\
&1.33/53&&&&&&&&&&&&&&&\\
J0746$^d$ &B:mek &  - & $2.2\pm_{0.7}^{1.0}$ & 48(6) & - & - & - & -  & $19\pm_{4}^{6}$ &  -  & - & $5.3(3)$  & 1 (fix)  & - & $8.9(4)$ & $\sim2.0$ \\
&0.89/52&&&&&&&&&&&&&&&\\
J0801 &brem+pow & $\leq0.016^a$ & $4.3(6)$ & $0.61(6)$ & - & - & - & $1.09(8)$  & 1.0(1)$^e$ & - & $1.2(2)$  & 0.14(4)$^e$ & - & $0.10(4)$ & 2.7(1) & $\sim4.8$ \\
&1.33/251&&&&&&&&&&&&&&&\\
J0927 &mek  & $0.06(1)$ & $4.7(4)$ & $84(1)$ & $33\pm^{4}_{3}$ & $76(2)$ & -     & -  & $12.6\pm^{0.9}_{1.3}$ & -       & - & $7.1\pm^{0.7}_{0.4}$   & $0.84\pm^{0.11}_{0.14}$ & $0.14(2)$ & $4.9(1)$ & $\sim2.2$ \\
&1.32/320&&&&&&&&&&&&&&&\\
J0958 &BB+mek  & $0.26(4)$ & - & - & $16.5(19)$ & $59(4)$ & $79(3)$     & -  & $36(14)$ &  $0.07\pm^{0.04}_{0.02}$      & - & $3.6(2)$   & $1.3\pm^{0.54}_{0.45}$ & $0.19(2)$ & $4.5(1)$ &  $\sim2.2$   \\
&1.08/294 &&&&&&&&&&&&&&&\\
J1701 &BB+2mek  & $0.76(4)$ & $7.7\pm^{0.7}_{0.4}$ & $65(1)$ & - & - & $53(1)$ & $6.0\pm^{1.1}_{0.4}$  & $\geq54^h$ &  $4.3\pm_{1.0}^{2.0}$  & $0.9\pm_{0.1}^{0.3}$ & $5.6(2)$   & $2.3(4)$ & $0.13(1)$ & $8.0(3)$ & $\sim41$  \\
&1.10/366$^f$&&&&&&&&&&&&&&&\\
J2113 &2mek  & $0.74(4)$ & $3.7(5)$ & $73(2)$ & $17(5)$ & $38(6)$ & - & $4.0(5)$ & $31(7)$ & - & $2.6(5)$ & $6.3(3)$  & $1.8(3)$ & $0.07(1)$ & $9.3(2)$ & $\sim2.6$ \\
&1.05/370$^f$ &&&&&&&&&&&&&&&\\
\hline
\end{tabular}  
\label{tab:averagespec}                      
\end{center}} 
\begin{flushleft}
$\alpha$ Blackbody normalization defined as $L_{39}/D^{2}_{10}$, where $L_{39}$ is the source luminosity in units of $10^{39}$ erg/s and $D_{10}$ is the distance to the source in units of 10 kpc. \\
$^a$ Multi-temperature power-law index $\alpha$ fixed to 1.\\
$^b$ $3\sigma$ upper limit (results refer to a model without this component).\\
$^c$ Fit results for EPIC data only.\\
$^d$ Fit results for XRT plus BAT data.\\
$^e$ Power law photon index and normalization.\\
$^f$ BAT inter-calibration constant is $0.50\pm0.07$ for J1701, and $0.60\pm0.12$ for J2113.\\
$^g$ $3\sigma$ lower limit.\\
\end{flushleft}                               
\end{table*}

\begin{table}
\caption{WD masses ($\rm M_{\rm WD}$), distances ($\rm d$), 
and mass accretion rates ($\rm \dot{M}$) of the sources in the sample.}
\begin{center} 
\begin{tabular}{lcccc}
\hline  
\noalign{\smallskip}
Source & $\rm M_{\rm WD}$ $^a$ & $\rm d$ $^b$ &  $\rm \dot M$ $^c$   \\
       & $\rm M_{\odot}$    &  pc    &  $\rm M_{\odot}\,yr^{-1}$   \\
\hline 
\noalign{\smallskip}
J0706      & $\geq0.79$ & 230  & $\sim1.3\times10^{-11}$   \\   
J0801      & $1.18\pm0.10$ & 250   & $\sim1.5\times10^{-11}$ \\
J0927      & $0.58\pm_{0.05}^{0.11}$ & $\lesssim670$  & $\lesssim2.2\times10^{-10}$   \\
J0958      & $0.74\pm_{0.12}^{0.11}$ & 720--1100   & $\sim 1.8-4.2\times10^{-10}$  \\
J1701      & $1.16\pm_{0.12}^{0.13}$ & 1000 & $\sim2\times10^{-9}$  \\
J2113      & $0.81\pm_{0.10}^{0.16}$ & $>750$ & $\gtrsim1.7\times10^{-10}$  \\
\hline
\end{tabular}
\label{tab:wdmass} 
\end{center}
$^{a}$ Derived from PSR model.\\ 
$^{b}$ Adopted distances. \\
$^{c}$ Derived from accretion luminosity (see text for details) and using the quoted masses. \\
\end{table} 

\begin{table*}
\caption{Spectral parameters at spin max and min. Other parameters are fixed to their average spectrum best-fit values. Uncertainties are at $1\sigma$ confidence level.}
{\small
\begin{center}
\tabcolsep=0.05cm
\begin{tabular}{cccccccccccccc}
\hline 
\\
source & N$_{\rm H_{ Pc1}}$ & cvf & N$_{\rm H_{ Pc2}}$ & cvf & kT$_{\rm BB}$ & kT$_{\rm c}$ & kT$_{\rm h}$ & n$_{\rm BB}$ & n$_{\rm c}$ & n$_{\rm h}$ & EW &  F$_{0.3-10}$ & $\chi^2$/dof    \\
       &  10$^{22}$ cm$^{-2}$ &  \%  & 10$^{22}$ cm$^{-2}$ & \%  & keV & keV & keV & $10^{-3}$ & $10^{-3}$ &  $10^{-3}$  & keV    & $10^{-12}$   &                     \\
       &                      &      &                     &     &     &     &     &           &           &             &        & \ergscm      &              \\                   
\hline 
J0706   &&&&&&\\
max ($\phi$=0.70-1.20)   & - & - &  $6.5\pm_{4}^{10}$ &  22(6)             & -  & -  & 45.8 (fix)  & - & - & 12.9$\pm$0.1 & - & 9.9(4) & 0.85/105 \\                                                                                                          
min ($\phi$=0.30-0.60)   & - & - &  $2.9\pm0.7$       &  $70\pm_{20}^{10}$  & -     & -   & $5.1\pm_{1.5}^{3.9}$  & - & - & 5.2$\pm$2  & - & $1.7\pm_{0.7}^{0.1}$ & 1.26/24 \\
\hline
J0801   &&&&&\\
max  ($\phi=0.10-0.40$)     & 4.6(7)               & 72(7)        &  -     &  -  & -    & 1.05(8)   &  0.8(3)$^c$     & - & 2.3(5) &  0.10(4)$^c$ & $<0.3$   & 3.1(1) & 1.09/137 \\
min  ($\phi=0.50-0.90$)     & $5.2\pm_{1.7}^{3.6}$ & 41(25)       &  -     &  -  & -    & 0.8(4)    &  1.4(3)$^c$     & - & 0.5(1) &   0.3(1)$^c$   &  0.3(1) & 2.4(2) & 1.13/141 \\ 
\hline
J0927   &&&&&&&\\
max ($\phi=0.91-1.10$)      & 2.8(5)  & 76(3) &  27(4)              &  74(3) & -   & -   & 17(5)            & - & - & 7.8(5) & 0.16(3) & 6.6(2) & 1.15/160 \\                                                                                                          
min ($\phi=0.71-0.85$)      & 6.2(17)  & 92(3) &  $32\pm_{7}^{14}$   &  76(7) & -   & -   & $18\pm_{4}^{13}$ & - & - & 6.8(8) & 0.16(5) & 4.7(3) & 1.03/47  \\
\hline
J0958   &&&&&&\\
max ($\phi=0.20-0.55$)      & -         & -     &  17.5(27)          &  61(3) & 0.078(4) & -   & $20\pm_{4}^{9}$      & 0.09(2) & - & 4.1(2) & 0.21(4) & 5.2(2) & 1.17/157 \\                                                                                                          
min ($\phi=0.75-0.99$)      & -         & -     &  15.0(25)          &  69(2) & 0.067(9) & -   & 36.27 (fix)$^a$      & 0.07(4) & - & 3.0(2) & 0.37(8) & 3.6(1) & 1.17/59 \\
\hline
J1701   &&&&&&\\
max  ($\phi=0.35-0.65$) &  7.4(7) & 52(2)  & - & - & 0.053(2) & $6.0\pm_{1.3}^{2.4}$ & 79.9 (fix)$^a$ & $5.1\pm_{1.2}^{1.6}$ & 0.90(3) & 6.6(4) & 0.11(2) & 9.6(1) & 1.16/271 \\
min  ($\phi=0.90-1.20$) & 10.2(7) & 82(1)  & - & - & 0.048(7) & 7.5(17)           & 79.9 (fix)$^a$ & $8.6\pm_{0.3}^{0.9}$ & 0.96(3) & 4.8(4) & 0.18(2) & 6.3(1) & 1.11/208 \\
\hline
J2113   &&&&&&\\
max  ($\phi=0.60-0.90$)   & 3.1(6)  & 67(4) &  15(4)              &  42(8)  & -   & 4(2)   & 28(6)  & - & 2.2(5) & 7.0(6)      & 0.07(2) & 10.1(3) & 0.88/256 \\
min  ($\phi=0.05-0.30$)   & 3.9(7)  & 78(4) &  $20\pm_{9}^{15}$   &  32(10) & -   & 5(2)   & $\geq22^b$ & - & 3.1(4) & 5.5(10) & 0.09(2) & 8.5(4) & 0.82/245 \\
\hline 
\end{tabular} 
\label{tab:pps_all}
\end{center}}
\begin{flushleft}
$^a$ Fixed to the average spectrum best-fit value.\\
$^b$ $3\sigma$ lower limit.\\
$^c$ Power law photon index and normalization.\\
\end{flushleft}
\end{table*}

\subsubsection{A hard X-ray and low-B field polar}
\label{subsub:mdot}

J0706 was identified in the nIR at $\rm K_s$=14.54 mag \citep{skinner14}.
It also appears in the 2MASS ca\-ta\-logue as 2MASS\,J07064892+0324472 with 
J=$15.585\pm0.065$, H$>$14.843 and K=$14.541\pm0.105$ mag. It is further detected in the mid-IR
with the Wide-field Infrared Survey Explorer
(WISE) \citep[][]{wright10} as WISE\,J07064895+0324469 in the W1 
(3.35$\rm \,\mu$m), W2 (4.6$\rm \,\mu$m), and W3 (11.6$\rm \,\mu$m) bands:
W1=$13.228\pm0.024$, W2=$12.563\pm0.026$ and W3=$10.751\pm0.116$ mag. 
The upper limit to the column density of a total absorber indicates that
extinction is negligible and translate to $\rm A_V\lesssim0.05$ 
\citep{Guver_Ozel09}. The 2MASS colour 
J-K=$1.04\pm0.12$ is consistent with a late type (M5--M6 V) donor star
\citep{Straizys2009}. A  M5.3 V spectral type 
is also expected for a donor in a 1.7 h binary \citep{Smith_Dhillon98}.
However, though WISE and 2MASS data are not contemporaneous, the K-W1 , W1-W2, and W2-W3 colours are much redder for such spectral
type \citep{pecaut13}, indicating a mid-IR excess. 
A number of polars studied  in the nIR and mid-IR have also shown 
extreme colours in the WISE bands \citep[W2-W3$\gtrsim$1;][]{Harrison13,bernardini14,harrison15}, 
ascribed to cyclotron radiation with
the fundamental harmonic falling in the W3 band \citep{harrison15}.
The IR excess in the case of cyclotron origin is also expected to be
variable at the orbital-rotational period, but the single exposure WISE
data are too sparse to reveal a modulation.
The position of J0706  in the mid-IR colour-colour diagram 
\citep[W1-W2;W2-W3, see Figure 1 in][]{harrison15} falls in a region consistent with
the presence of a cyclotron fundamental in the W3 band. This suggests a moderately
low magnetic field ($\sim7-10\times10^{6}$ G). 
We then use 2MASS K-band magnitude to estimate the distance.
Adopting a M5--M5.6 V donor with 
$\rm M_K=8.3-8.6$ \citep{knigge06}, the observed K band magnitude 
implies a distance of 150--300 pc. We therefore assume a distance of 230 pc
for J0706. 
An estimate of the mass accretion rate is obtained
assuming for the accretion luminosity $\rm L_{acc} =G\,\dot M\,M_{WD}\,R_{WD}^{-1}
\sim \,L_{X,bol} + L_{cyc}$,
where $\rm  L_{X,bol} $ and $\rm L_{cyc}$ are the 
X-ray optically thin bolometric and cyclotron luminosities.
The latter, as estimated from the mid-IR flux excess, is $\rm L_{cyc} 
\sim 3.3\times10^{30}\,erg\,s^{-1}$ for a distance of 230 pc, and contributes
only $\sim 2\%$ to the total luminosity. With $\rm L_{acc} 
= 1.2\times\,10^{32}\,erg\,s^{-1}$, adopting 
$\rm M_{WD}=0.79\, M_{\odot}$ 
and its corresponding radius of 7.07$\times10^{8}$ cm \citep{nauenberg72}\footnote{The WD 
Mass-Radius relation by \cite{nauenberg72} is adopted
throughout this work.}, a mass accretion rate of 
$\rm \dot M \sim 1.3 \times 10^{-11}\,M_{\odot}\,yr^{-1}$ is obtained (Table \ref{tab:wdmass}). This  
is in remarkable agreement with the expected secular mass transfer rate for
gravitational losses \citep{Warner95} for a system with $\rm P_{orb}$=1.7 h 
and a mass ratio q=0.25 having adopted $\rm M_2=0.2\,M_{\odot}$ for the M5 donor
\citep{knigge06}. 

The lack of a soft X-ray blackbody component 
is not surprising given the recent discovery of an increasing number of polars 
not displaying such soft X-ray emission \citep{ramsay04b}. 
A shift to the EUV or FUV bands due to large footprints may be possible. However, as pointed out by \cite{ramsay09}, there is no obvious
explanation for why these hard polars should have larger accretion areas, 
since their orbital period and magnetic field 
strengths are not so different from other polars. 
The hard X-ray detection of polars also poses the question of whether 
the few (now 11) that are hard X-ray selected possess lower magnetic fields 
and/or more  massive WDs. J0706 appears to show both properties.

\subsection{Swift\,J0746.3-1608}
\label{J0746}

J0746 was identified as a CV independently by \cite{thorstensen13}, where it figures as Swift\,J0746.2-1611, and 
\cite{parisi14}, and proposed as either a novalike system or a magnetic system because of its optical spectral characteristics. 
From the late-type secondary absorption lines \cite{thorstensen13} 
derived a long orbital period of 9.38 h and a distance of 
$900\pm^{190}_{150}$ pc. 

\subsubsection{Timing analysis}

\begin{figure*}
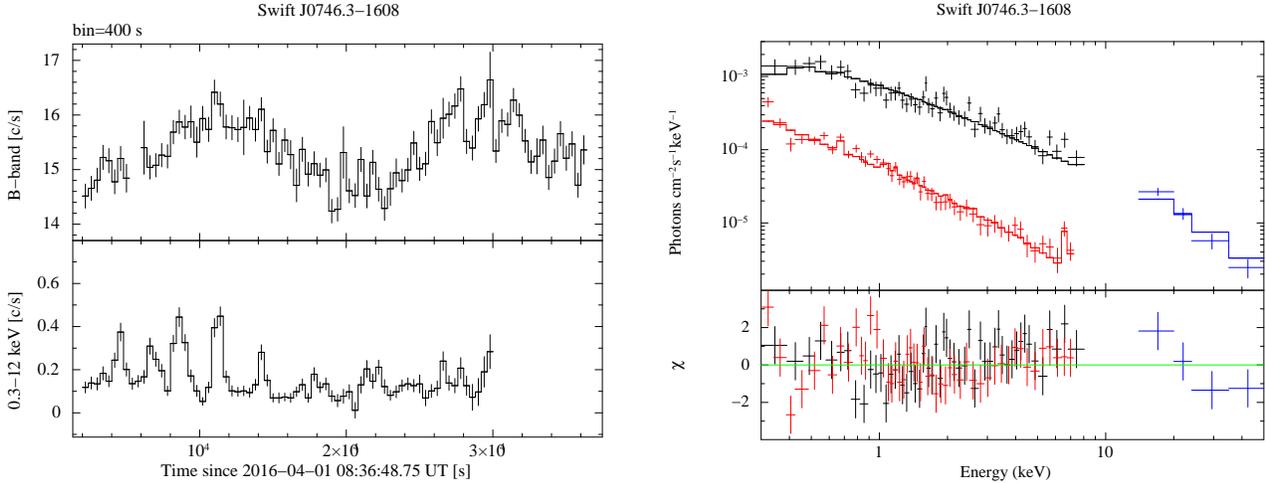

\begin{center}
\begin{tabular}{cc}
\includegraphics[angle=270,width=3.3in]{j0746_lc_all.eps} & \includegraphics[angle=270,width=3.3in]{j0746ufs.eps}\\ 
\end{tabular}
\caption{\textit{Left:} Optical OM B-band (upper panel) and X-ray PN 0.3-13 keV (lower panel) 
light curves of J0746. The B-band light curve is modulated at $\sim$5 h, about 
half the 
orbital period, while the X-ray light curve shows flaring in the first part of 
the pointing. \textit{Right:} Broadband 
unfolded spectrum of J0746. Post fit residuals are shown in the \textit{lower panel}. 
In black are \Swift/XRT data (0.3--10 keV), in blue \Swift/BAT data (15--50 keV), and 
in red \XMM/PN (0.3--10 keV) data. The fit uses model A in Table 
\ref{tab:averagespec} and it is made separately on \Swift\ (XRT+BAT) and 
\XMM\ (EPIC) data.
The spectra are plotted together to highlight the source flux variability.}
\label{fig:j0746}
\end{center}
\end{figure*}

The X-ray lightcurve\footnote{The 
last quarter of the \XMM\ pointing was removed from the analysis because it was affected by high background level.} shows a faint source ($\sim0.14$ c/s, in the PN). In the 
first half of the pointing there are at least five flare-like events of variable 
amplitude (up to a factor of 6) with timescale of about 2700 s (Figure 
\ref{fig:j0746}), but these do not appear to be periodic. 
The HR vs time are constant within uncertainty. The optical 
(B-band) light curve (Figure 
\ref{fig:j0746}), where no flares are detected, shows instead  
a  modulation  with amplitude $\rm \Delta$ B = 0.10 mag. Although 
covered only for two cycles, we estimate a period of 5.03$\pm$0.10 h, 
about half the spectroscopic orbital period found by 
\cite{thorstensen13}. These authors did not detect any 
modulation in their photometry of Jan. 2013, but instead an erratic variability. This indicates that this source is variable on a long-term timescale (yrs).  

Indeed \cite{thorstensen13} presented the \Swift-XRT light curve of J0746 covering
the period between 2009 and 2011. The source is highly variable  with count
rate changing up to a factor of $\sim20$ on timescale of a few hours. 
Two additional pointings in Jun.\,2013 and Aug.\,2015 
showed the source had faded to a count rate of $\sim$0.05 c/s 
from a maximum of about $\sim0.7$ c/s during 2009--2011. 
It then appears that a substantial change of the X-ray fluxes occurred
after 2011. We then inspected the \Swift-UVOT photometry
acquired since 2009, in the U, UVW1 and UVW2 filters. The source
also faded in the UV and U bands after June 2011, by $\sim$1.3 and $\sim$1.0 mag, respectively.
We further inspected the optical long-term history using  B-band 
photometric measures from USNO-A2 catalogue (B=13.4 mag, in 1953), USNO-B1 catalogue (B1=14.5
and B2=14.7 mag, in 1971), the Guide Star Catalogue 
(GSC\,2.2) (B=$14.55\pm0.41$ mag in 1984), and the APASS catalogue (B=$15.02\pm0.02$ mag in Jan.2011). 
These, compared with the OM photometry (B=$16.33\pm0.03$ mag), further
confirm the fading at optical  wavelengths. The double humped 
orbital modulation in the B band observed with 
the OM could be due to ellipsoidal modulation of the donor star filling its Roche lobe. The two minima
of the light curve correspond to the inferior and superior conjunction of the
secondary, according to the spectroscopic ephemeris of \cite{thorstensen13}.
The lack of multi-band photometry does not allow to further investigate the origin of the optical modulation and so to confirm this hypothesis.

\subsubsection{Spectral analysis}

Due to the remarkable long-term X-ray variability, the EPIC spectra were
fitted without the high-energy coverage from the \Swift-BAT instrument, which
would require an unfeasible inter-calibration constant as high as 100. This suggests 
that when observed by \XMM\ the source was much fainter than its average over many years. 
The spectrum (Table \ref{tab:averagespec} and Figure \ref{fig:j0746}) is satisfactorily fitted with a  \textsc{mekal} with kT$=7.4\pm0.7$ keV 
and A$_{\rm Z}=2.1\pm0.4$ ($\chi^2=1.1$, 125 d.o.f.). Similar
$\chi^2$ is also obtained using a simple power law (with photon index $1.70\pm0.05$), but an
excess results at the iron complex. This makes us prefer the former model.
The absorption is very low, with a $3\sigma$ upper limit of
N$_{\rm H}\leq3\times10^{20}$ cm$^{-2}$, at least a factor $\sim$10 
lower than the total galactic column density in the
direction of the source that lies in the dust lane and consistent with the 
optical extinction by \cite{thorstensen13}. 
The absorbed 0.3--10 keV flux is low ($\rm 4.6\times 10^{-13}\,
erg\,cm^{-2}\,s^{-1}$) indicating a very low mass accretion rate and consequently a negligible intrinsic
absorption. Furthermore, the lack of a fluorescent line at 6.4 keV is also
consistent with such low state, where reflection from cool material 
is negligible. No significant spectral change, except for the normalization of the \textsc{mekal},
is found extracting the in-flare and out-flare spectra, as also indicated by the constant HR.

Given the source long-term X-ray variability, we  analysed the average 
XRT spectrum 
which covers the higher state observed between Aug.\,2009-Jun.\,2011 (IDs: 38960, 
40698, 41163 and 90159).  
We fitted the broad-band XRT+BAT spectrum adopting a thermal component, since 
a power law fails to account for the BAT data. We first used model A, a \textsc{mekal} 
(with the abundance fixed to Solar) 
multiplied by \textsc{phabs} obtaining kT$=39\pm_{11}^{24}$ keV and 
N$_{\rm H}=6\pm2\times10^{20}$ cm$^{-2}$ ($\chi^2=1.33$, 53 d.o.f.).  
An improved fit is found using model B, where a partial covering absorber is used in place of 
\textsc{phabs} (when using both, the latter is totally unconstrained) and gives kT$=19\pm_{4}^{6}$ keV, N$_{\rm H_{Pc}}=2.2\pm_{0.7}^{1.0}\times10^{22}$ cm$^{-2}$, and cvf=$48\pm6$ per cent ($\chi^2=0.89$, 
52 d.o.f.). Here we note that the former fit gives a column density in line
with the estimated extinction \citep{thorstensen13}. 
We notice that for both models the BAT inter-calibration constant is now consistent 
with unity, within uncertainty.
The XRT 0.3--10 keV flux is $\sim 8.3-8.9\times 10^{-12}\,\rm erg\,cm^{-2}\,s^{-1}$, 
a factor $\sim$17 larger than in the \XMM\ observation (Table \ref{tab:averagespec} 
and Figure \ref{fig:j0746}). 
If instead compared with the maximum flux recorded by XRT, 
the source in April 2016 is a factor of $\sim70$ fainter.

\subsubsection{An atypical low accretion rate CV or a LMXB?}

J0746 faded after 2011, and in April 2016 it was 
at its lowest level ever observed. 
At the distance of 900 pc, the bolometric X-ray luminosities in the 
high and low states are $\rm 2.0\times 10^{33}\,erg\,s^{-1}$ and
$\rm 5.8\times 10^{31}\,erg\,s^{-1}$, respectively. 
Both values, for a WD mass in the range 
0.6--0.8 M$_{\odot}$ would give accretion rates of 
$\rm \sim 2.0-3.4\times 10^{-10}\,M_{\odot}\,yr^{-1}$ and 
$\rm \sim 0.6-1.1\times 10^{-11}\,M_{\odot}\,yr^{-1}$, respectively.
 These  fall too short (at least 2--3 orders of magnitudes) 
with respect to those of systems above the gap 
\citep{McDermott_Taam89,howell01}, unless evolved donors
are taken into account \citep{goliasch_nelson15}. 

The short and long-term X-ray and optical variabilities observed in 
J0746 are difficult to reconcile with a long-period nova-like CV that is expected
to have a large accretion disc that acts as a reservoir.
Furthermore the lack of periodic pulsations (with the caveat of the low S/N) 
in the typical range of MCVs, also rules out the possibility that this systems harbours a magnetic WD. The peculiar flaring activity, the low luminosity level in both high and low
states poses the question on whether J0746 is indeed a CV or a low-mass
X-ray binary (LMXB). 
X-ray variability and luminosities down to $\rm 10^{31}-10^{33}\,erg\,s^{-1}$ are observed in transient LMXBs (with both neutron star or black hole primaries) during quiescence \citep[e.g.][]{hynes04,degenaar12,bernardini13a,bernardini14a}. 
Those harbouring neutron stars (NS) 
show X-ray spectra generally revealing
the optically thick thermal emission from the NS atmosphere at several tens of eV 
and/or a power law hard tail ascribed
to non-thermal emission processes possibly related to the magnetic field of the NS
\citep[see e.g.][]{degenaar11,degenaar12}. Black hole LMXBs, so far discovered
through their strong outbursts except for a recent detection by \cite{tetarenko16},
show hard X-ray power laws during quiescence 
\citep[with power law index very close to 2][e.g. higher than in the case of 
J0746]{plotkin13}.
The \Swift\ long-term monitoring does not give evidence that J0746 has 
undergone an outburst over the last 7 years and its spectrum, especially during the high state, seems to be thermal (and not a simple power law).
On the other hand, the observed behaviour in J0746 
may be reminiscent of the sub-luminous states of two LMXBs, PSR\,J1023+0038 and XSS\,J12270-4859, harbouring 
millisecond pulsars (MSPs) that initially were wrongly
classified as CVs \citep[see][]{archibald09,demartino10}.
These pulsars were observed to transit from LMXB to
MSP radio phases and vice versa 
and consequently defined as transitional MSPs \citep[tMSPs;][]{archibald09,demartino10,demartino13,bassa14,
demartino14,demartino15,bogdanov15}.  
During the LMXB state, they show peculiar high (persistent emission) 
and low (dips) modes  besides erratic flaring activity, with msec pulses 
detected during the high mode only \citep{archibald15,papitto15}. 
In J0746 dipping behaviour is not observed, but we are unable to 
search for pulsations in the 1--10 msec range, due to the temporal resolution of the 
EPIC camera.
Furthermore, differently from J0746, the X-ray spectra of tMSPs are featureless and described with
a simple power law. 

Additionally, at the X-ray levels observed in J0746 and drawing similarities with
two tMSPs, a detection of a Gamma-ray source by {\it Fermi}-LAT would have been possible.
However, we inspected possible Gamma-ray positional associations of J0746 and found
that the only close {\it Fermi}-LAT source is 3FGL\,J0748.0-1639, but it is 37.6\,
arcmins away. The $\rm 4^{th}$ {\it Fermi}-LAT  catalogue 
\citep{acero15} gives a 95$\%$ error ellipse of 
0.12\,deg$\times$0.14\,deg, excluding a positional association with J0746. Furthermore, this source has a different Gamma-ray 
spectrum than those of the tMSPs. We also inspected radio catalogues through the {\it Heasarc} website and found no radio source consistent with the position of J0746 \footnote{The closest source seems to be OI-173, a bright (200 mJy at 20 cm) radio source, but is 2.9\,arcmin away from J0746 \citep{dixoncat}.}.
In summary, the lack of positional Gamma-ray or radio association does not support a MSP classification.

While a quiescent LMXB in  a prolonged (years) low state could not be excluded, 
the thermal spectrum of this source may favour an atypical low luminosity CV. 
High and low states seen in J0746 could be explained if the
donor is not completely filling its Roche-lobe allowing for fast changes in
the accretion rate as well as accounting for the low state after 2011.
CVs entering extended low states are seen in the VY\,Scl star 
subclass, but they
have shorter orbital periods (typically 3--4 h). This system poses the same
questions regarding the true nature of the low-luminosity CV,  V405\,Peg, 
which is however just at the upper end of the VY\,Scl stars orbital periods 
\citep[$\rm P_{orb}=4.2$ h;][]{schwope14}. The other type of CVs undergoing
high and low states on timescales of years are the polars, but our data do 
not support a magnetic nature. In particular at the long orbital 
period of J0746, no Polars are known except for V1309\,Ori 
($\rm P_{orb}=8.0$ h, B$=4.5\times10^{7}$ G), which however is a blobby-accretor and displays one of the most
extreme soft-to-hard flux ratios \citep{demartino98,schwarz05}. J0746 does not share any
similarity with this system. On the other end, the
optical spectra acquired in 2010 \citep{thorstensen13,parisi14} show intense 
emission lines  of Balmer series and He\,II,  with ${\rm H_{\beta}}$ 
and He II(4686 \AA) equivalent width ratios similar to those observed in 
polars and  novalikes  and much larger than 
those observed in LMXBs \citep{vanparadis84}.

In conclusion there are no strong indications favouring either an 
LMXB or a typical long-period CV, but if its X-ray spectrum is indeed  thermal as it seems more likely, we suggest it could be a CV accreting at an unusually low rate for its
long orbital period. This is not in disagreement with the 
recent results by \citep{pala17} who also found several long-period CVs
accreting at unexpectedly low rates.  
Both an X-ray monitoring and a new long X-ray observation of J0746 
in a higher state are desirable to shed light into the true nature of
this peculiar source.

\subsection{PBC\,J0801.2-4625}
\label{J0801}

J0801, =1RXS J080114.6-462324 in \cite{masetti10}, was identified by those authors as a CV and tentatively suggested 
as magnetic. 
It was observed in April 2016 by \Swift-XRT due to a possible, though 
uncertain, association
with the fast soft X-ray transient MAXI J0758-456 \citep{masumitsu16,kennea16}. 

\subsubsection{Timing analysis}

The X-ray and V-band light curves display clear short term variability.
The power spectra in both bands show a strong, isolated, peak at 
$\sim7.6\times10^{-4}$ Hz. We measure a period of $1310.9\pm1.5$ s in the X-ray band 
and $1306.3\pm0.9$ s in the V-band, respectively that we interpreted as
the spin period of the accreting object. The 0.3--12 keV and V-band pulses 
are broadly
in phase and have similar PF ($15.3\pm1.5$ per cent  and $13.8\pm0.3$ per cent, respectively). 
The X-ray PF slightly decreases with energy, from $20\pm2$ per cent in the softer 
to $9\pm6$ per cent in the harder band (Table \ref{tab:pf} and Figure \ref{fig:pulseprof}).

Given the purported association to the transient detected by MAXI a compact star spinning at 1300 s could be either a NS or a WD.
However, slow NS rotators at the rate observed in J0801 are found only 
in high-mass X-ray binaries, and in the few known symbiotic X-ray binaries  
 and none detected in LMXBs \citep{patrunowatts12}.
The optical spectrum \citep{masetti10} clearly disfavors an 
early type counterpart as well as a K--M giant.
Furthermore, the presence of the pulsation in both the X-ray and optical bands is typically
observed in CVs being the optical spin modulation produced in the
magnetically confined accretion flow onto the WD poles. The more energetic
LMXBs do not generally show optical spin pulsations.
Therefore, the X-ray and optical modulations in  
J0801 strongly suggest a MCV of the IP type. 
The presence of only the spin period further 
implies that that accretion occurs via a disc \citep{rosen88}.

\subsubsection{Spectral analysis}

The EPIC spectra of J0801 are  not easy to reconcile with the
typical X-ray spectra of magnetic CVs. This is the only source of 
the sample for which a model based on a thermal continuum do not provide the best fit in terms of $\chi^2$. The combined \XMM\ and \INT\ broad-band spectrum can instead be described
by a simple model  (Table \ref{tab:averagespec} and Figure \ref{fig:avrspec}) consisting of \textsc{bremss} with kT$_{c}=0.98\pm0.06$ keV
and a power law with photon index $1.0\pm0.1$ (dominating for $E\gtrsim3$ keV)
plus a Gaussian at 6.4 keV with EW$=100\pm40$ eV, all multiplied by
\textsc{pcfabs} with N$_{\rm H_{Pc}}=4.0\pm0.6\times10^{22}$ cm$^{-2}$, 
and cvf=0.61$\pm$6 per cent. 
Only a $3\sigma$ upper limit to the  hydrogen 
column density of the total absorber is obtained 
(N$_{\rm H}\leq1.6\times10^{20}$ cm$^{-2}$), a factor of about 15 
lower than the total value in the source direction.
The hard component is not compatible with a second \textsc{bremss} since
an unrealistic temperature of 200 keV is obtained, even  
including the effect of reflection from neutral material 
(\textsc{refl} in \textsc{Xspec}). 

On top of its peculiar spectrum, we also note that $\chi^2$ is slightly high (1.33, 252 d.o.f.), 
and we found that this is in part due to post-fit residual features at low energies ($\lesssim3$ keV). Due to the low S/N, they could be modeled with either absorption or emission lines (we note that in the latter case a \textsc{mekal} thermal continuum with solar abundance could fit the data). Unfortunately, the low X-ray flux of J0801 prevents us using the RGS spectra to study in details the soft portion of the X-ray spectrum. Consequently, we decided not to attempt to model these features and consider the adopted average spectrum a satisfactory description of the data. 
Due to these complexities, the PSR model has been applied to the broad-band spectrum above 3 keV from which a massive WD ($\rm M_{WD}1.18\pm0.10\rm\,M_{\odot}$) is 
obtained (Table \ref{tab:wdmass}).  

Only the normalization of the thermal component increases by a factor of $\sim$5 at pulse maximum, while
all other parameters are constant within uncertainty (Table \ref{tab:pps_all}).
The spin modulation then seems mainly due to visibility changes of the
post-shock region, rather than to absorption effects.

\subsubsection{A low accretion rate IP?}

Due to the source faintness during the \XMM\ pointing 
and the purported association with the fast MAXI transient source, 
all available archival \Swift-XRT observations 
(obsid 34515 in 2007 and 262347 in 2016) were inspected for long-term changes.
Similar fluxes are measured during these observations, indicating a likely
intrinsically faint source. 
This does not exclude the possible association with the bright 
(1.4 Crab) fast-flaring source, and thus that J0801 could be a LMXB in
deep quiescence. However, while the X-ray spectrum is atypical for a
CV, the presence of pulses in both X-ray and
optical bands strongly supports J0801 harbouring a magnetic WD. 
Furthermore, the optical spectrum by \citep{masetti10} acquired in 2009
displays strong emission
lines (Balmer, He\,I and He\,II) with EW ratio of 
$\rm H_{\beta}$ and He\,II close to those observed in magnetic CVs and
much larger than those of LMXBs \citep{vanparadis84}.

J0801 is detected in the nIR as 2MASS\,J08011702-4623274, with K=$14.442\pm0.041$, 
H=$14.032\pm0.050$, and K=$13.774\pm0.058$ mag, as well as in the midIR as 
WISE\,J080116.96-462327.5 with W1=$13.307\pm0.025$, W2=$13.153\pm0.027$, and W3=$12.401\pm0.311$ mag.
As shown in Sect.\,4.3.2, extinction is negligible. The nIR  
colour J-H=$0.41\pm0.06$ suggests a K-type donor star, but those at longer wavelengths, 
H-K=$0.28\pm0.08$ and W1-W2=$0.15\pm0.04$ are instead both compatible 
with a M4 V star \citep{bilir08,Straizys2009,pecaut13}. This implies that accretion affects the emission above 1.2$\mu$.
For an assumed M4\,V donor a distance of 250 pc is obtained using the
observed and absoluted K-band magnitudes ($\rm M_K=6.76$) \citep{knigge06}. 
Such small distance would imply the closest LMXB ever known, especially if associated to the fast MAXI transient source. All this
makes the magnetic CV interpretation more plausible.
  
The lack of multi-colour photometry does not allow to infer a possible
contribution of reprocessed emission at UV/optical wavelengths as would be
expected from the detection of pulsations in the V-band \citep[see 
also][]{mukai94}. However given the small optical pulse 
amplitude ($\sim14$ per cent), reprocessing should not be important. 
 We then estimate the accretion luminosity adopting a distance of 250\,pc as: 
$\rm L_{acc} \sim L_{X,bol}=3.6\times 10^{32}\,erg\,s^{-1}$. 
Using the inferred WD mass (Table \ref{tab:wdmass}) 
a mass accretion rate 
$\rm \dot{M}\sim 1.5\times 10^{-11}\,M_{\odot}\,yr^{-1}$ is derived.
This value is close to that of a system loosing angular momentum through 
gravitational radiation below the 2--3 h orbital period gap. 
J0801 may resemble those short period 
IPs, like HT Cam \citep{demartino05}, with a spin pulse not 
affected by absorption and  accreting at a low rate. 
Although the binary period cannot be measured with the present data, 
if the donor spectral type is indeed M4 V, J0801 is expected to have 
$\rm P_{\Omega}\sim2-3$ h  \citep[][]{knigge06}, 
quite consistent with the above arguments.
Time-resolved  optical spectroscopy is needed to assess 
whether J0801 is indeed a short orbital period system.
Also, the peculiar X-ray spectrum requires a more detailed analysis with
higher quality X-ray spectroscopy to classify it as an "ironclad" IP.

\subsection{Swift\,J0927.7-6945}
\label{J0927}

J0927 was identified from optical spectroscopic follow-ups by \cite{parisi14}, as a CV, likely of the magnetic type. It appears as PBC J0927.8-6945 in their work.

\subsubsection{Timing analysis}

The X-ray and optical light curves display short-term and long-term variability (Figure \ref{fig:orbital}).
The X-ray power spectrum shows a strong peak at $\sim9.7\times10^{-4}$ Hz and
weaker ones at twice and three times this frequency. Power is also detected
close to the main peak at $\sim9.2\times10^{-4}$ Hz. 
We measure $\rm P_1=1033.54\pm0.51$ s 
and $\rm P_2=1093.4\pm6.5$ s, with an amplitude ratio 
A$^X_{1}$/A$^X_{2}\sim2.5$. The shorter period is then identified 
as the WD spin period and the other as the beat ($\omega-\Omega$). 
The power spectrum also shows substantial power at low frequencies. The 
lightcurve indeed displays almost two cycles of a possible periodic modulation, which 
could be the orbital period.
It is asymmetric and we use two sinusoid
functions (fundamental and harmonic) to determine its period:
P$^{X,lc}_{\Omega}=5.15\pm0.10$ h. 
This is consistent with that obtained from the beat frequency ($\omega - \Omega$): 
P$^{X,side}_{\Omega}=5.25\pm0.45$ h. Thus, J0927 is undoubtedly an IP. 
The presence of a dominant spin pulsation in the X-ray band implies that it mainly accretes 
through a disc, although a non negligible fraction ($\sim40$ per cent) of the 
accretion flow overpasses the disc and impacts directly onto the WD poles. 
The optical B band light curve instead shows evidence of the spin period 
only ($\rm P^{opt}_{\omega}$=$1030.6\pm0.9$ s), together with a 
longer timescale (hours) non periodic variability.

The presence of harmonics up to the second implies a structured X-ray spin
pulse profile (Figure \ref{fig:pulseprof}). The two main peaks giving rise to
the fundamental and first harmonic are separated by 0.4 in phase. The PF decreases
with increasing energy (Table \ref{tab:pf}), but the first harmonic almost 
dominates the 0.3--1 keV energy range (PF up to $\sim 70$ per cent), and  
loses power as the energy increases. This may indicate the presence
of two emitting poles with the lower one dominating at low energies.     
The optical pulse profile is instead almost sinusoidal ($\Delta B\sim 0.1$ mag),
with a single broader maximum approximately encompassing the phase interval covered 
by both X-ray peaks, 
and centred on the first harmonic X-ray peak (Figure \ref{fig:pulseprof}). 
This suggests that the optical and X-ray emitting regions are somehow linked 
and that the former is likely more extended. 

The HR shows hardening at orbital minimum and the intensity 
of the orbital modulation decreases with energy. These findings imply 
that as for the spin pulse, also the X-ray orbital modulation
is mainly due to localized absorption.

\subsubsection{Spectral analysis}

We fitted the average broad-band spectrum (Table \ref{tab:averagespec} and Figure 
\ref{fig:avrspec}) with a single \textsc{mekal} (kT$=12.6\pm_{1.3}^{0.9}$ keV) plus a 
Gaussian (EW=$140\pm20$ eV), multiplied by \textsc{phabs} 
(N$_{\rm H_{Ph}}=6\pm1\times10^{20}$ cm$^{-2}$) and 
two \textsc{pcfabs} (N$_{\rm H_{Pc1}}=4.7\pm0.4\times10^{22}$ and N$_{\rm H_{Pc2}}=3.3
\pm_{0.3}^{0.4}\times10^{23}$ cm$^{-2}$). We got a slightly high $\chi^2$ (1.3 for 294 d.o.f.), 
also because of the residuals below 1.5 keV. Unfortunately, data from the RGS cannot help in 
resolving residual features, since the source is heavily absorbed.
The relatively low temperature of the emitting plasma, although being an average
over the post-shock region,  may indicate that the WD is
not massive. Indeed, the PSR model gives a WD mass of 
$\rm 0.58\pm^{0.11}_{0.05}\,M_{\odot}$ (Table \ref{tab:wdmass}).

The spectrum at pulse minimum is more absorbed. The Pc1 parameters increase  
by a factor of $\sim$2.2 (N$_{\rm H_{Pc1}}$) and of $\sim$1.2 ($\rm cvf_{Pc1}$).
We conclude this is the main cause of the X-ray pulse. The high column density absorber 
(Pc2) is in fact constant within uncertainty (Table \ref{tab:pps_all}).  
The spectrum at orbital minimum is also affected by absorption. However, if both Pc1 and Pc2 parameters
are left free to vary, Pc1 is still  the dominant component 
(Table \ref{tab:pps_orb}). Due to the low S/N we are not able to
infer spectral changes along the spin period at orbital maximum and minimum.

Energy dependent X-ray orbital modulations have been observed in a number of IPs \citep{parker05,bernardini12} 
indicating that material fixed in the binary frame, such as the stream impact point at the outer disc rim, could produce 
additional absorption. Since J0927 show evidence of (a partial) disc-overflow accretion mode, this interpretation seems well justified.  

\begin{table}
\caption{Spectral parameters at maximum (Max) and minimum (Min) of the orbital modulation (Orb.). Other parameters are fixed to their average spectrum best-fit values. Uncertainties are at $1\sigma$ confidence level.}
{\small
\begin{center}
\tabcolsep=0.05cm
\begin{tabular}{cccccccc}
\hline 
\\
source & Orb.  &  N$_{\rm H_{ Pc1}}$    & cvf  & N$_{\rm H_{ Pc2}}$   & cvf & F$_{0.3-10}$ & $\chi^2$/dof    \\
       &       &  10$^{22}$ cm$^{-2}$  &  \%  & 10$^{22}$ cm$^{-2}$ & \%  &   $10^{-12}$ &   \\
       &       &                       &      &                     &     &   \ergscm    &   \\ 
\hline  
J0927 & Max  & 3.8(3)  & 79(1) &  29(1) &  74(2) & 5.4(1)  & 1.34/296 \\                                                                                                          
      & Min  & 6.9(7)  & 93(1) &  38(2) &  78(3) & 4.2(1)  & 1.14/172  \\
\hline
J2113 & Max  & 3.3(5)  & 64(4) &  12(2) &  37(7)  & 9.9(1)  & 0.97/344 \\
      & Min  & 4.4(4)  & 89(1) &  30(4) &  47(4)  & 7.7(2)  & 0.86/225 \\
\hline 
\end{tabular} 
\label{tab:pps_orb}
\end{center}}
\end{table}

\subsubsection{An IP above the gap}

J0927 shows temporal and spectral characteristics of a typical IP.
A tentative distance of 240 pc was derived adopting 
an absolute magnitude
of $\rm M_V \sim9$ \citep{parisi14}. We here try to constrain the source distance as follows.
The low column density  of the total absorber suggests low extinction.
Using the relation by \cite{Guver_Ozel09} an upper limit to
the interstellar extinction is derived:
$\rm A_V \lesssim0.27$. 
J0927 is also identified as 2MASS J09275308-6944
with J=$14.823\pm0.042$, H=$14.642\pm0.062$ and K=$14.311\pm0.084$ mag, and as 
WISE J092752.94-694442.3, with W1=$14.273\pm0.059$, W2=$14.005\pm0.052$ and 
W3=13.459$\pm$0.42 mag. The dereddened nIR colours 
J-H=$0.15\pm0.07$ and H-K=0.32$\pm$0.1, indicate the presence of 
more than one spectral component. The J-H colour would 
suggest a mid F-type star, but the H-K index would give a spectral type 
M2--M6 \citep{Straizys2009}. The mid-IR colour W1-W2=0.27$\pm$0.08 also 
indicates a spectral type later than M4 \citep{pecaut13}. 
A M2.3 V donor is expected in a 5.15 h binary 
\citep{Smith_Dhillon98,knigge06}.  
Since  later spectral type than M4  would imply a
system below the gap, we 
 assume a donor in the range M2--M4 \citep[hence absolute
magnitudes $\rm M_K=5.16-7.69$,][]{knigge06} and use the 
dereddened K-band magnitude
to estimate the distance that results in the range 
$\sim210-670$ pc. While this encompass the value found by 
\cite{parisi14}, it poses
a more stringent upper limit to the distance.

We then evaluate the accretion luminosity and then accretion rate in 
J0927. Also in this case the contribution of reprocessed radiation cannot be
determined with the present data. Using the upper limit of 670 pc,  
$\rm L_{acc} \sim L_{X,bol} \lesssim 1.2 \times 10^{33}\,erg\,s^{-1}$, 
which in turn translates into a mass accretion rate $\rm \dot M  
\lesssim 2.2\times 10^{-10}\,M_{\odot}\,yr^{-1}$ for a 0.58$\rm M_{\odot}$
WD mass (Table \ref{tab:wdmass}).
This is $\sim2$ orders of 
magnitude lower than the secular mass transfer rate expected for a 
binary evolving through magnetic braking in 
a 5 h orbit \citep{McDermott_Taam89,howell01}. Lower 
mass transfer rates than predicted by the present-day CV population models 
were also recently found in a number 
of CVs above the gap \citep[see][]{pala17}.
While diversities in their secular properties are expected for selected
CV populations \citep{goliasch_nelson15}, any conclusion needs to await 
accurate {\it GAIA} parallaxes.

The spin-orbit period ratio is $\rm P_\omega/P_\Omega\sim$0.06, as it is
for most of the IPs observed above the orbital period gap and
accreting via a disc truncated at the magnetospheric radius \citep[see][]{norton08}.  
For a WD spinning at equilibrium, the corotation radius
($\rm R_{co}=(G\,M_{WD}\,P_{\omega}^2/4\,\pi^2)^{1/3}$) defined as the radius at which the magnetosphere rotates 
at the expected Keplerian frequency for that radius, balances 
the Alfven radius, $\rm R_{A}=5.5\times10^8\,(M_{WD}/M_{\odot})^{1/7}\,R_9^{-2/7}\,
L_{33}^{-2/7}\,\mu_{30}^{4/7}\, cm$. Here $\mu$ is the magnetic moment in units of
$10^{30}\rm G\,cm^{-3}$, $\rm L_{33}$ is the luminosity in units of 
$\rm 10^{33}\,erg\,s^{-1}$, and the WD mass and radius are in solar and 
$10^9$ cm units, respectively.  
Adopting P$_{\omega}$=1033.54 s, the luminosity determined above, 
${\rm M_{WD}=0.58\,M_{\odot}}$ and $\rm R_{WD}=8.8\times 10^8$ cm,
a magnetic moment 
$\rm \mu \sim 5\times 10^{32}\,G\,cm^{-3}$ is inferred.
This is at the low end of magnetic moments expected for
slow rotators  \citep{norton99,norton04}, as it is the WD in J0927, but
it depends on the assumed accretion luminosity.

J0927 appears a typical IP above the period gap and caught 
in an hybrid accreting mode, which is not uncommon in long period systems
\citep[see][]{norton97,beardmore98,bernardini12}.

\subsection{Swift\,J0958.0-4208}
\label{J0958}

J0958 was identified by \cite{masetti13} as a CV, possibly of the magnetic type through optical spectroscopy. 
  
\subsubsection{Timing analysis}

The X-ray power spectrum shows a strong peak corresponding to a period of
$296.22\pm0.05$ s that we interpret as the WD spin period and a much weaker feature at its first harmonic 
(Table \ref{tab:time}). There is no sign of these short-term 
periodicities in the optical (B-band) power spectrum.
The inspection of the X-ray light curve suggests that long-term X-ray variability (on a timescale of $\sim8$ h) could be present, although  
consistent with the length  of the pointing. A possible variability 
on a timescale of  $\sim2.2$ h may be present in the optical band, but does not
appear to be periodic. J0958 is also a variable source 
(CSS\,J095750.7-420836) 
monitored in the Catalina (CRTS)\footnote{http://crts.caltech.edu/}  survey from 2006 to 2013, 
but a search for periodicities less than a day did not give conclusive  
results. Thus, the orbital period of this system remains unknown.

The PF decreases as the energy increases (Table \ref{tab:pf} and Figure 
\ref{fig:pulseprof}) from a maximum of about 38 per cent in the 0.3--1 keV band 
to a minimum of about 7 per cent in the 5--12 keV 
band\footnote{The PF is calculated 
removing the last 12 ks of EPIC exposure which are affected by high background activity}. 
The HR shows hardening at spin minimum, as found in the majority IP 
systems.
J0958 seems to be an IP dominated by its spin period and therefore a pure disc accretor.

\subsubsection{Spectral analysis}

The best fit model ($\chi^2=1.08$, 294 d.o.f.) to the average broadband spectrum 
(Table \ref{tab:averagespec} and Figure \ref{fig:avrspec}) consists of 
a \textsc{mekal} (kT$=36\pm13$ keV) plus a blackbody (kT$=79\pm3$ eV) 
and a Gaussian (EW$=190\pm20$ eV), absorbed by a total \textsc{phabs} (N$_{\rm H}=2.6\pm0.4\times10^{21}$ cm$^{-2}$) and a partial
covering absorber \textsc{pcfabs} (N$_{\rm H_{Pc}}=16.5\pm1.9\times10^{23}$ cm$^{-2}$). 
Metal abundances are consistent with solar value. No additional optically thin 
component is required. 
The blackbody component is hot, but not unusual as found also in other soft IPs
\citep[see][]{haberl02,anzolin08,bernardini12}. 
The column density of the total absorber 
is within 2$\sigma$ consistent with that in the direction of
the source \citep[$\rm N_{H,gal}=1.6\times10^{21}\,cm^{-2}$,][]{kalberla05}.
The fit with the PSR model gives  
$\rm M_{WD} 0.74\pm_{0.12}^{0.11}\,M_{\odot}$ (Table \ref{tab:wdmass}), 
which is consistent within uncertainties with that resulting when adopting a 
36 keV maximum temperature ($\sim$0.8$\rm M_{\odot}$).
The main changes in the spectral parameters with the spin period
are related to variations in the normalization of the
\textsc{mekal}, which increases at spin maximum by a factor of $\sim$1.4. 
The cvf of \textsc{pcfabs} also slightly increases at pulse minimum ($69\pm2$ vs 
$61\pm3$ per cent), but not its column density. 
The other parameters remain constant within uncertainties (Table \ref{tab:pps_all}). The lack of 
variability of the blackbody normalization indicates that the irradiated polar cap of the
WD remains visible along the whole spin pulse.

The bolometric soft-to-hard flux ratio defined as $\rm F_{soft}/4\,F_{hard}$ 
\citep[see][]{ramsay04b} is 0.09, similar to those found in other soft
IPs with comparable hot blackbody temperature \citep{anzolin08,bernardini12}. 
The reprocessed X-ray emission would then originate from a small polar
region as inferred from the normalization of the blackbody: 
$\rm A_X \sim 0.65 - 3.4 \times 10^{13}\,cm^{2}$ for a distance in 
the range 0.72--1.1 kpc (see below).
Indeed, adopting a $\rm M_{WD}=0.74\rm\,M_{\odot}$ and its corresponding $7.5\times10^{8}$ cm radius, a fractional area $f \sim
0.1-6\times 10^{-6}$ is obtained.

\subsubsection{A fast spinning IP}

J0958 shows typical IP characteristics. A tentative distance of 200 pc was derived by \cite{masetti13} using similar arguments as in J0927 by \citep{parisi14}. We here estimate the distance using nIR and mid-IR photometry. Both the high column density derived from X-ray spectral fits and the interstellar reddening in the direction of the source \citep{Schlafly11} should be considered as upper limits to the galactic extinction to the source since the optical spectrum \citep{masetti13} does not reveal interstellar absorption features (DIBs). We then assume A$_{\rm V}<0.65$.

J0958 is detected as 2MASS J09575064-4208355 with J=$15.420\pm0.067$, 
H=$14.926\pm0.075$ 
and K=$14.865\pm0.118$ mag and as WISE J095750.67-420835.8 with 
W1=$14.455\pm0.030$ and W2=14.148$\pm$0.042 mag.
The J-H=0.49$\pm$0.10 colour suggests a $\sim$K1--K5 V donor star . A similar range (K0--K4) is obtained using the dereddened colour 
\citep{Straizys2009}, while W1-W2=0.27$\pm$0.05 
points to a spectral type later than M4 \citep{pecaut13}. However, a fit to the combined 
nIR and mid-IR SED 
requires only one spectral component and gives a blackbody temperature 
of $4.1\pm0.4$ kK, indicating a K4--M2 star.
Using the more accurate J-band magnitude,  
assuming a K4--M2 V donor star with $\rm M_J=5.18-5.97$ mag 
\citep{bilir08,knigge06}, a distance 
of 0.72--1.0 kpc (0.77--1.1\,kpc) is obtained with (and without) reddening 
correction. We conservatively adopt the distance range 0.72--1.1 kpc. 

The lack of pulsations in the optical band may suggest that additional X-ray 
reprocessing is negligible. We then estimate the accretion luminosity including
both soft and hard X-ray emission: $\rm L_{acc}\sim L_{X,thin}+L_{X,soft}$,
which using the above distance range results in: $\rm \sim 1.4 - 3.4 
\times 10^{33}\,erg\,s^{-1}$.
For a 0.74$\rm M_{\odot}$ WD (Table \ref{tab:wdmass}) the
accretion rate is 
$\rm \dot{M} \sim 1.8 - 4.2 \times10^{-10}\,M_{\odot}\,yr^{-1}$. 
This is rather similar to what found in long orbital period 
IPs (e.g. above the period gap). If the donor spectral type is 
truly K4--M2, the binary period would be in fact be above the gap 
($\rm P_{\Omega} \sim5-6$ h). 

J0958 hosts a fast spinning WD only surpassed by a few systems: SWIFT\,J0525.6+2416 (226 s), 
XY\,Ari (206 s), 
V2731\,Oph (128 s), DQ\,Her (71 or 142 s), AE\,Aqr (33 s), and V455\,And
(67 s) \citep[see][for a recent update of IPs]{ferrario15}. Except for the last example, 
they are all long ($\gtrsim4$ h) orbital period systems with 
spin-to-orbital period ratios $\lesssim$ {0.01. While the lack of knowledge
of the orbital period prevents us from determining the degree of asynchronism,
the above arguments may favour a system above the gap and 
thus J0958 may in future also join this group.
Systems with such very small ratios are expected to be disc accretors 
when the WD is spinning at 
equilibrium, and would be strong propellers if out of equilibrium, as the 
unique system AE\,Aqr \citep{norton08}. The detection of X-ray spin pulses  
then shows that J0958 is a disc accretor IP that possess a
WD spinning at equilibrium. Furthermore, fast rotating
WDs are expected to harbour relatively weak magnetic WDs, as opposed to
slow rotators with strong magnetic fields \citep{norton99}. Thus, J0958 could
have a magnetic moment as low as $\rm \lesssim 10^{32}\,G\,cm^{-3}$.

\subsection{Swift\,J1701.3-4304}
\label{J1701}

J1701, =IGR J17014-4306 in \cite{masetti13}, was identified by those authors as a CV and proposed to be a magnetic candidate
for its optical spectroscopic characteristics.  

\subsubsection{Timing analysis}
\subsubsection*{The eclipse}

All the three EPIC light curves show at the middle of the observation a 
remarkable drop of count rate that reaches zero, indicative of a total 
eclipse of the X-ray source (Figure 
\ref{fig:orbital}). A partial eclipse is also detected in the simultaneous OM B-band
photometry, where the flux drops by a factor of $\sim3$. An archival \XMM\ observation (obsid: 0743200101) 
of the planetary nebula PN\,G343.3-00.6=PN\,HaTr\,5, which is only 
17\,arcsec apart from J1701 was carried out
on 2014-08-31 for 14\,ks.
J1701 is serendipitously detected in the EPIC exposures
as a bright source overwhelming PN\,G343.3-00.6, which is instead not detected.
At the beginning of this pointing, the final part of a total X-ray eclipse clearly appears.
This shows that this feature is real and that J1701 is a deep eclipsing CV.
The lack of consecutive eclipses in both \XMM\ observations implies an
orbital period $\gtrsim4.4$ h.

The ground-based optical and nIR photometry (Section 2.4) 
also displays eclipses in all bands. 
The {\it LCO} photometry caught one full eclipse and another 
only partially. Four eclipses are instead detected in the
{\it REM} data, and seven in the {\it AAVSO} photometry.
To determine the orbital period the combined g' and V-band, 
light curves were analysed with the \textsc{Period04} 
package\footnote{http://www.univie.ac.at/tops/Period04/}. The power
spectrum reveals a strong peak at 1.87 d$^{-1}$ and period analysis gives 
$\rm P_{\Omega}=0.53413\pm0.00002$ d, and epoch of minimum $\rm T_{0}$(BJD)=2457660.53569$\pm$0.00003. 
An improved ephemeris was obtained extending the baseline 
with the LCO g'-band (June 2016) and the OM B-band data (2015). Each of the
13 eclipses was fitted with a Gaussian to obtain the times of minima. 
A linear regression  was used to obtain the following ephemeris: 
$\rm T_{0}(BJD)=2457599.66277\pm0.00009$ and $\rm P_{\Omega}=0.534026\pm0.000005$ d.
Observed-minus-calculated ($O-C$) residuals were inspected against trends, 
showing that a constant period gives an acceptable fit. The  excursions on 
average are within 240 s around zero. 
The eclipse depths derived from Gaussian fits are: $\Delta g'=0.96\pm0.03$, $\Delta r'=0.72\pm0.04$, $\Delta i'=0.59\pm0.03$, 
$\Delta B=1.27\pm0.07$, $\Delta V=0.68\pm0.02$ and $\Delta J=0.30\pm0.08$ mag. The width
of the optical eclipses (\textsc{fwhm}) are instead consistent within errors ($2765\pm90$ s).
The X-ray eclipse lasts $2920\pm30$ s and is total for $2747\pm15$ s. 
The egress appears slightly asymmetric since it happens at the peak of a spin pulsation (Figure \ref{fig:orbital}).
To measure the e-folding times we fitted the light curve (14 s binned) outside the eclipse with two sinusoids to account for the X-ray spin pulse, 
and with an exponential decay/rise to model the ingress/egress time and found $\rm \tau_{ingress}=94\pm42$ s and $\rm \tau_{egress}=63\pm14$ s (Figure \ref{fig:lc_optical}).

\begin{figure}
\includegraphics[width=3.5in, height=4.5in, clip=true, trim=0 5 0 5]{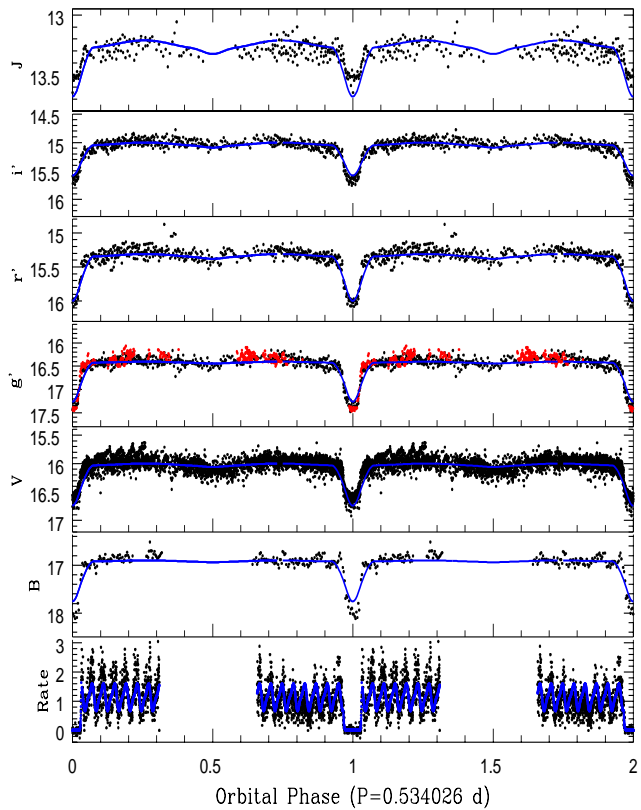} 
\caption{From bottom to top: The X-ray and optical light curves of J1701 folded 
at the 12.8 h period using the ephemeris quoted in the text. 
For the g' band the LCO photometry is denoted with 
red points. The blue lines report the {\sc Nightfall} curves for the optical photometry (see text for details), 
and for the PN X-ray light curve, where a model with two sinusoids describing the spin and its first harmonic, plus two exponential
functions describing the eclipse ingress and egress, were used.}
\label{fig:lc_optical}
\end{figure}

\subsubsection*{The spin pulsation}

The X-ray persistent emission further displays a periodic modulation. The PN
power spectrum has a strong peak at $\sim5.5\times10^{-4}$ Hz and its first 
harmonic. No other signal is detected. By removing the eclipse interval, 
a composite two-sinusoidal fit gives: P$^X_{\omega}=1859\pm3$ s (Table \ref{tab:time}). 

The OM B-band light curve also displays similar periodic variability. A sinusoidal fit
does not require the first harmonic and gives P$^{opt}_{\omega}=1857\pm6$ s, fully
consistent with the X-ray period. We used the {\it AAVSO} V-band light curve to improve the measure of the spin period. A fit with two-sines to model the fundamental and its first harmonic 
gives P$^{opt}_{\omega}=1858.67\pm0.02$ s. This period is adopted to fold and study the
X-ray pulse properties (Table \ref{tab:time}). Both the X-ray and optical (B and V bands)
pulses are in phase indicating a common origin. The X-ray PF decreases from 70 per cent 
 in the 0.3--1 keV band to 18 per cent in 5--12 keV band 
(Figure \ref{fig:pulseprof} and Table \ref{tab:pf}), suggesting that photoelectric absorption is the main cause of the X-ray spin pulsation.

\subsubsection{Spectral analysis}
\label{j1701spec}

The best fit to the average broad-band spectrum (Table \ref{tab:averagespec} 
and Figure \ref{fig:avrspec}) consists of two \textsc{mekal} (kT$_{\rm c}=6.0\pm_{0.4}^{1.1}$ keV and kT$_{\rm h}\geq54$ keV, $3\sigma$ lower limit), plus a blackbody 
(kT$_{\rm BB}=53\pm1$ eV) and a 6.4 keV Gaussian (EW=$110\pm30$ eV). The hydrogen column density  
of the total absorber is N$_{\rm H}=7.6\pm0.4\times10^{21}$ cm$^{-2}$ and that
of the partial is  N$_{\rm H_{Pc}}=7.7\pm_{0.4}^{0.7}\times10^{22}$ cm$^{-2}$ with ${\rm cvf=65\pm1}$ per cent ($\chi^{2}_{\nu}=1.1$ for 366 d.o.f). The metal abundance is high (A$_{\rm Z}=2.3\pm0.4$), 
though consistent with solar values within $\sim3\sigma$. 
The BAT inter-calibration constant is $0.50\pm0.07$, suggesting that the sources could 
be variable on the long-term and that when observed by \XMM , it was brighter than the 
average over many years. Despite its hard spectrum, a reflection component is not 
required. Furthermore, the PSR model gives a massive WD ($1.16\pm_{0.12}^{0.13}\,M_{\odot}$, 
Table \ref{tab:wdmass}). This high value is consistent with the lower limit obtained 
for the hot \textsc{mekal} component ($\rm M_{WD}\geq0.97\,M_{\odot}$). 

The HR shows an hardening at spin minimum. The equivalent hydrogen column 
and the covering fraction of the partial covering absorber both increase at 
spin minimum by a factor of $\sim1.4$ and $\sim1.6$, respectively. 
The normalization of the hot mekal slightly increases by a factor of about 1.4 at maximum. 
On the other hand, the blackbody normalization 
increases by a factor $\sim 1.7$ at spin minimum, indicating this component 
arises from the heated WD atmosphere at the poles. Similarly, the EW of the 
6.4 keV fluorescent line also increases from 110 eV at spin maximum 
to 180 eV at spin minimum, suggesting that reflection occurs at the WD poles (Table \ref{tab:pps_all}).

We derive a bolometric flux ratio $\rm F_{soft}/4\,F_{hard}\sim1.95$, which is the highest ever inferred in a soft X-ray IP and closer to that observed in polars \citep{anzolin08,bernardini12}. 
Although subject to model uncertainties, the emitting area of the WD hot spot is 
$\rm 4.1-7.8 \times 10^{15}\,cm^2$ at a distance of 1 kpc (see below).
This would correspond to a large fractional area, $f \sim 1.5-3.9 \times 10^{-3}$, of a WD with $\rm M_{WD}\sim1.1\,M_{\odot}$. 

\subsubsection{A deep eclipsing long period IP}

The multi-band photometry has been analysed to unveil 
the donor star and system parameters. Using the B, V, g'
r', i' and J bands magnitudes (Figure \ref{fig:lc_optical}), 
the spectral energy distributions outside and at eclipse minima 
were constructed. 
We applied a reddening A$_V$=3.17 derived assuming that the \textsc{phabs} absorber, 
which 1$\sigma$ lower error is N$_{\rm H}=7.2\times10^{21}$ cm$^{-2}$, is entirely
interstellar and then applying the extinction relation by \cite{Guver_Ozel09}. 
Although this source lies in a heavily reddened region of the Galactic Plane,
a higher extinction is disfavoured by inspection of the optical spectrum
by \cite{masetti13}, where weak DIBs are identified. Fitting the SED at eclipse minimum
with a blackbody gives a temperature of $7.3\pm0.3$ kK indicating a relatively
hot donor. 
The optical spectrum presented in \cite{masetti13} also reveals  the {\it Mg\,b} 
feature at 5167-5184\,\AA , 
which was compared with stellar templates from the Pickles catalogue 
\citep{pickles98}
with spectral types  in the range F0--G0 V using several reddening values. 
The best match is found for a  F0--F5 V spectral type with A$_V$=3.1, in 
agreement with the SED at eclipse minimum, with the mid-F spectral type being preferred. 
The contribution of the donor star to the total V-band 
flux outside the eclipse is $\sim53$ per cent.

While a detailed analysis and fine tuning of the optical/nIR light curves
will be presented in a forthcoming work, first estimates  of the binary system parameters, 
the mass ratio $q$, the inclination $i$, and the donor star temperature
were obtained with the {\sc Nightfall} code (v.1.88) \footnote{http://www.hs.uni-hamburg.de/DE/Ins/Per/Wichmann/Nightfall.html} 
applied simultaneously to the most covered orbital light curves in V, g', r', i' and J bands. 
These were fitted adopting a Roche-lobe filling donor star and a primary
treated as a point source with fixed temperature of 200 kK (e.g. the higher 
allowed temperature). An irradiated
accretion disc with fixed inner and outer radii of 0.01 and 0.40 in units of
the WD Roche-lobe was also included. The disc thickness was set 
to 1 per cent and the temperature
of the inner disc boundary and of the hot-spot at the outer disc rim to 30 kK 
and 8 kK, respectively\footnote{Details of parameters setting can be found in the
{\sc Nightfall} User Manual}.

The donor temperature, $q$, and $i$ were left free to vary obtaining
$q=0.87\pm^{0.22}_{0.20}$, $i=72\pm^{1}_{2}$ deg and $\rm T_2=7.2\pm0.4$ kK
($\chi_{\nu}^2$=28 for 5640 dof). 
The large $\chi_{\nu}^2$ is due to the large scatter
and small errorbars (typically 3 per cent) of the photometric measures and to
the approximate modeling of the eclipse feature. 
Due to the extent of the wings of the eclipse a larger accretion disc is disfavoured \citep[see also][]{horne82}. 
Adopting ${\rm M_{WD}=1.16\pm_{0.12}^{0.13}\,M_{\odot}}$, the donor 
would have  $\rm 0.7 \lesssim\, M_2 \lesssim\, 1.4\,M_{\odot}$. 

The length of the X-ray eclipse, and particularly the time elapsed from
the first to the $\rm 3^{rd}$ contact, (or the $\rm 2^{nd}$ and $\rm 4^{th}$ 
contact) 
$\rm \Delta \phi_{orb}\simeq $0.06, allows the placing of
constraints on the donor star radius in units of binary separation: 
$\rm R_2/a = (sin^2\,(\pi\,\Delta \phi) + cos^2\,(\pi\,\Delta \phi)\,\times 
cos^2\,i)^{1/2}$  \citep{horne82}. Assuming a Roche-lobe filling secondary,
this equation,  together with the equivalent Roche lobe  radius 
approximation by \cite{eggleton83}, define an unique relation between $q$ and $i$.
Adopting the $1\sigma$ uncertainty on $q$, $i$ and 
$\rm M_{WD}$ as a range, this restricts the orbital separation to 
$\rm a \sim 3.33 - 3.87\, \rm R_{\odot}$  and the donor radius to
$\rm R_2 \sim 1.14-1.48\, \rm R_{\odot}$. The secondary appears to be
expanded and evolved, in line with the expected present-day 
population  of CVs with evolved donors 
at these long orbital periods \citep{goliasch_nelson15}.
Adopting the range of the donor radius and using the normalization obtained
from the fits to the SED during the eclipse, a distance of 995--1290 pc is obtained.
We then adopt a distance of 1 kpc for J1701.

Furthermore, an estimate of the X-ray emitting region $\rm R_x = (v_1 + v_2)/2 \, \Delta T_{egress}$ 
is also derived. Here $\Delta T_{egress}=63\pm14$ s and $\rm v_1$ and $\rm v_2$ 
are the primary and secondary relative velocities. From Kepler's third law: 
$\rm v_1 + v_2 = (2\,\pi\,G (M_1 + M_2)/P_{\Omega})^{1/3} \, sin\,i$. 
Adopting the $1\sigma$ uncertainties on
the WD and donor masses, inclination and egress time,
we obtain 
$\rm R_x \sim 0.7 - 1.3 \times 10^9\,cm$, which is $\sim1.3-3.4$ 
times the radius of the massive WD. 
Taken at its face value, this would suggest a very extended X-ray 
emitting region. Post-shock regions are only a few percent the WD  
radius \citep[see][]{mukai99}.
To account for such large linear extension, two opposite emitting poles should be visible
at the same time, as suggested by
the presence of the harmonic of the spin frequency, and in turn they should be gradually hidden during the eclipse ingress/egress.

Adopting  for the  accretion luminosity 
$\rm L_{acc} \sim L_{X,thin}+L_{X,soft}$,
at a distance of 1 kpc it results in 
$\rm \sim 4.9 \times 10^{34}\,erg\,s^{-1}$. For a $\rm M_{WD}=1.16\,M_{\odot}$, we derive  $\rm \dot M \sim 2 \times 10^{-9}\,M_{\odot}\,yr^{-1}$ 
(Table \ref{tab:wdmass}). 
At long ($\sim8-10$ h) orbital periods, 
donors in CVs are expected to have high mass transfer rates and to be 
chemically evolved \citep{goliasch_nelson15}. The inclusion
of nuclear evolution in the synthesis models of the present-day CV population 
allows the attainment of longer ($\gtrsim10$ h) periods and a wide range of 
mass-transfer rates, including the value inferred here for J1701. 
The probability of finding such very long period systems is however
low, about 1 per cent. 

J1701 is one of the longest orbital period IP. It is surpassed only by
GK\,Per (47.9 h) and V2731\,Oph (15.4 h). These are highly 
asynchronous systems ($P_{\omega}/P_{\Omega} \sim 0.002$), while
J1701 has $P_{\omega}/P_{\Omega}$ = 0.04.
Accounting for the high mass ratio inferred, the observed asynchronism in J1701
is consistent with a WD spinning at equilibrium \citep[][]{Wynn_King95,King_Wynn99} and accreting via a disc \citep{norton08}. We then derive  
a magnetic moment $\rm \mu \sim 5 \times 10^{33}\,G\,cm^{-3}$.
Though to be regarded with caution, this large value would suggest that
J1701 will evolve into a Polar when reaching synchronism
\citep{norton04}. 

J1701 is then the first deeply eclipsing IP ever found at long orbital periods and
adds to the small group of eclipsing systems of this class: 
EX\,Hya, XY\,Ari, IPHAS\,J062746+0148, V597\,Pup, and DQ\,Her 
\citep[][and references therein]{hellier14}, and the two recently discovered IPs
 SWIFT\,J201424.9+152930 \citep{esposito15} and CXOGBS\,J174954.5-294335
\citep{johnson17}.  In addition, IGR\,J18293-1213 \citep{clavel16} 
is definitely an eclipsing CV and may well be an eclipsing IP.
Time-resolved optical/nIR polarimetry 
will be crucial to understand the evolution of this IP and 
the interaction of the accretion flow with the magnetosphere.

\subsection{Swift\,J2113.5+5422}
\label{J2113}

J2113 was tentatively proposed by \cite{masetti10}, as a magnetic CV of the IP type 
given the strength of its He II emission line. It appears as 1RXS\,J211336.1+542226 in their work.

\subsubsection{Timing Analysis}

The X-ray power spectrum clearly shows three peaks, pointing to three periodic signals. 
A broad peak at low frequency $6.9\times10^{-5}$ Hz, together with two close peaks at 
$7.3\times10^{-4}$ Hz and $7.9\times10^{-4}$ Hz. 
We then identify the 
highest frequency peak as the WD spin and the close one as 
the beat ($\omega-\Omega$), while the 
broader low-frequency peak as the binary frequency ($\Omega$).  
Indeed, the X-ray light curve fully samples three orbital cycles 
(Figure \ref{fig:orbital}) 
with P$^{X,lc}_{\Omega}=4.02\pm0.10$ h. 
We further measure $\rm P_{\omega-\Omega}=1373.8\pm2.6$ s and 
$\rm P_{\omega}= 1265.6\pm4.5$ s (Table \ref{tab:time}). 
Using the beat and spin frequency a period 
$\rm P_{\Omega}^{X,Side}=4.46\pm0.10$ h is obtained. It is longer
than P$^{X,lc}_{\Omega}$, but consistent within 3$\sigma$. 
The OM V-band light curve instead does not show any short-term
periodicity, but rather large flickering (Figure \ref{fig:orbital}). It is modulated at a period 
$\rm P_{\Omega}^{opt}=3.63\pm0.11$ h, which is shorter, but 
consistent within  3$\sigma$, with the X-ray value. 
The amplitude of the optical modulation is smaller than that in 
the X-ray band ($16\pm2$ vs $29\pm1$ per cent; Figure \ref{fig:orbital}). 
The intensity of the spin modulation in the whole 0.3--12 keV range is low (PF$\sim15$ per cent),
but we still measure a higher PF at lower energies (ranging from 30 per cent
in the softest range to 9 per cent in the hard band. 
(Table \ref{tab:pf} and Figure \ref{fig:pulseprof}). 
The HR varies along the orbital period, with the source being 
harder at the 
orbital minimum. The intensity of the orbital modulation also decreases with 
energy. All these findings suggest that photoelectric absorption is the 
main cause of both the spin and orbital modulations. 
As in the case of J0927, the fact that the orbital X-ray modulation is energy dependent suggests that 
the X-ray emission is absorbed by material fixed in the binary frame, such as the outer disc rim. 
The amplitude ratio of the two short-period X-ray modulations is 
A$^X_{\omega}$/A$^X_{Side}\sim1.1$, indicating that the system is in a
disc over-flow accretion mode.

\subsubsection{Spectral analysis}

We fitted the average broad-band spectrum (Table \ref{tab:averagespec} and 
Figure \ref{fig:avrspec}) with two \textsc{mekal}
(kT$_{\rm c}=4.0\pm0.5$ and kT$_{\rm h}=31\pm7$ keV) plus a Gaussian, 
absorbed by a total \textsc{phabs} (N$_{\rm H,Ph}=7.4\pm0.4\times10^{21}$ cm$^{-2}$) and 
two partial covering columns \textsc{pcfabs} (N$_{\rm H_{Pc1}}=3.7\pm0.5\times10^{22}$ and N$_{\rm H_{Pc2}}
=1.7\pm0.5\times10^{23}$ cm$^{-2}$) ($\chi^{2}_{\nu}=1.05$ for 370 d.o.f.). We note 
that the BAT inter-calibration constant is $0.60\pm0.12$, which may suggest that the 
source is variable on the long-term and that we observed it in a state brighter than 
the average over many years. The EW of the 6.4 keV fluorescent 
line is very small (70 eV), indicating that reflection is not important
in the hard X-ray portion of the spectrum.
The column density of the total absorber ($\rm N_{H,Ph}$) is consistent
with the galactic hydrogen column density in the source direction  
\citep[$\rm N_{gal}=8.5\times10^{21}\,cm^{-2}$;][]{kalberla05}. 
The PSR model gives a WD mass of $0.81\pm^{0.16}_{0.10}\,\rm M_{\odot}$, 
consistent within uncertainties with that resulting from the 
hot \textsc{mekal} component (Table \ref{tab:averagespec}). 

Also because of the weak spin modulation, the pulse 
resolved spectroscopy only shows marginally significant changes. The spectrum is 
slightly harder at pulse minimum, where $\rm cvf_1$, the covering fraction  
of the lower density absorber (Pc1), increases by a factor of $\sim$1.2 (Table \ref{tab:pps_all}). 
On the other hand, the spectrum at orbital minimum is clearly  
harder, mainly due to the increase by a factor $\sim$2.5 of the
hydrogen column density of the second partial covering absorber (Pc2) and 
a to a slight increase of $\rm cvf_2$, its covering fraction 
(see Table \ref{tab:pps_orb}). Therefore, this higher density absorber is
mainly responsible for the orbital modulation. The spectral analysis 
then clearly shows that two different absorbers are responsible 
for the spin (Pc1) and the orbital (Pc2) modulation. They can
be ascribed to the magnetically confined pre-shock flow and matter accumulating
at the disc rim, respectively.

\subsubsection{An IP above the period gap}

J2113 lies close to the galactic plane (b$\sim4^o$) and thus
interstellar extinction in the direction of the source is high.
We then adopt the column density of the total absorber
($\rm N_{H,Ph}=7.4\times10^{21}$ cm$^{-2}$) as an upper limit to the interstellar extinction in the source direction, which translates into $\rm A_V$=3.3 \citep{Guver_Ozel09}. Also for J2113 we 
estimate the distance using nIR and mid-IR data.
However, 
J2113 is not a 2MASS source, but surprisingly it appears as 
WISE\,J211335.41+542233.0 with W1=$14.985\pm0.038$ and W2=$14.519\pm0.048$ mag.
The dereddened colour W1-W2=$0.43\pm0.06$, would
indicate a very low temperature component $\sim1800$ K. 
For a 4.1 h orbital period binary, the donor should have a spectral type
of a M3.4 V star \citep{knigge06}, and thus much bluer 
colours (W1-W2$\sim$0.12).
This points to an excess of flux, as also found in some  
CVs. It could be due to circumbinary dust emission or, as seen in 
J0706 to low-harmonic cyclotron flux in the case of polars.
Further analysis is beyond the scope of this paper, but investigation in 
the IR is needed to unveil the true nature of the excess in this system.
From the lack of detection in the 2MASS survey, we estimate an 
upper limit to the K-band flux (lower limit magnitude of 
$\rm K > 15.4$) in a region close to the position of the source in the 2MASS survey.
This, corrected for interstellar extinction, translates 
into a lower limit to the distance of 750 pc,  
adopting a M3.6 V star with $\rm M_K=6.3$ \citep{knigge06}.
A rough $\sim1.6$ kpc distance would instead result using the 
extinction-distance relation as a function of galactic latitude. 

The accretion luminosity is then evaluated as: 
 $\rm L_{acc} \sim L_{X,bol} \gtrsim  1.7\times 10^{33}\,erg\,s^{-1}$
adopting conservatively d$>$750\,pc. For a WD mass of 0.81$\rm M_{\odot}$
(Table \ref{tab:wdmass}), this translates into 
$\rm \dot M \gtrsim 1.7\times10^{-10}\,M_{\odot}\,yr^{-1}$. As a lower
limit it could be consistent with those of long period systems \citep{howell01}.
} 

J2113 is then an IP above the gap with a typical  
spin-to-orbit period ratio of $\sim0.09$. 
For a WD spinning at equilibrium 
a magnetic moment $\rm \mu \gtrsim 6\times10^{32}\,G\,cm^{-3}$ is derived,
consistent with a moderately magnetic slow WD rotator.  
This system is then another example of a disc-overflow accretor above
the orbital gap.   

\section{Conclusions}
\label{sec:conclusions}

The broad-band analysis presented here unambiguously reveals the magnetic 
nature of five hard X-ray selected CV candidates.
We indeed classify J0927, J0958, J1701, and J2113 as firm IPs. We suggest that J0801 is also an IP, although possibly displaying an atypical 
spectrum. J0706 is instead a polar, joining as the $\rm 11^{th}$ member
the group of hard X-ray selected polars.
One system, J0746, is found to accrete at an extremely low rate for its 
long orbital period (9.38 h). The lack of periodic variations in its X-ray light curve does not support
the magnetic CV membership. It could be either an atypical low-luminosity CV
or even a LMXB, although evidences for this interpretation are not 
compelling.

The class of IPs has grown substantially over the last
decade \citep[see e.g.][]{ferrario15},   
now amounting to 66 systems \citep[with 10 of them identified 
in the last two years,][]{bernardini15,esposito15,cotizelati16,tomsick16,johnson17}. 
This is about half of the size of the Polar class ($\sim110$ members).

\begin{figure*}
\begin{center}
\begin{tabular}{cc}
\includegraphics[angle=0,width=3.3in]{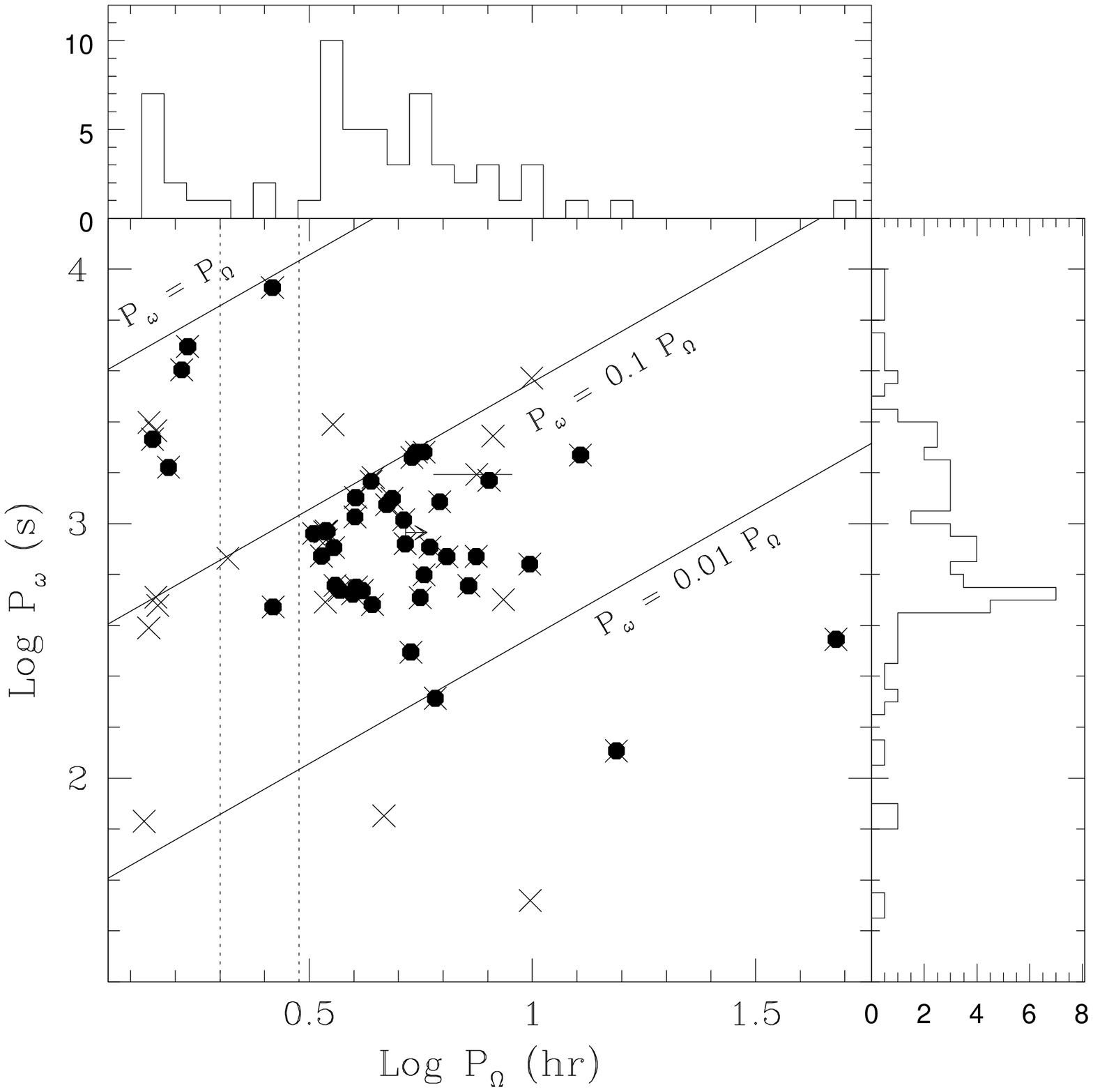} & 
\includegraphics[angle=0,width=3.3in]{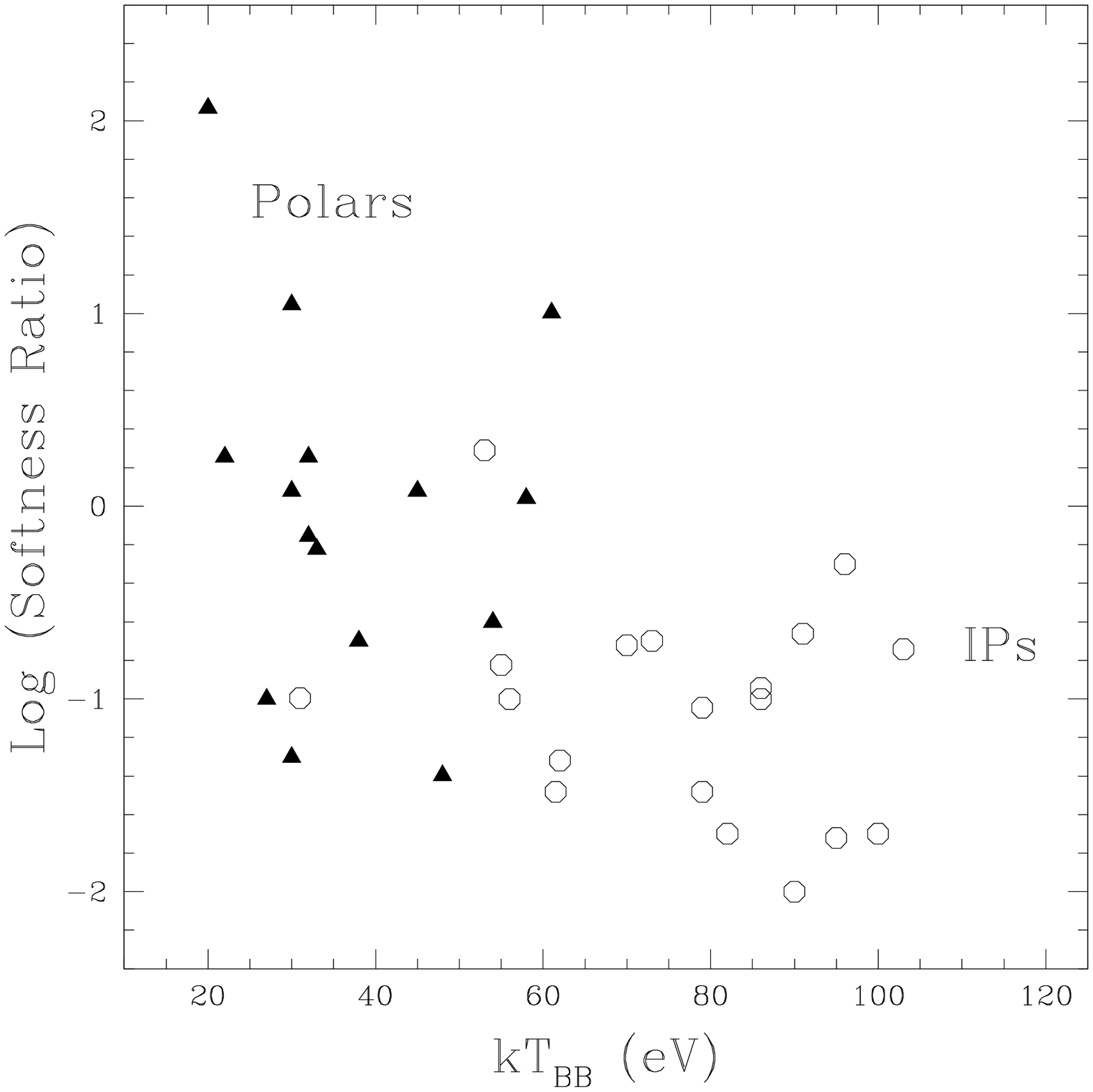}\\ 
\end{tabular}
\caption[]
{{\it Left:} The spin-orbit period plane of confirmed IPs (crosses) with
those identified in the hard \Swift\ and \INT\ surveys denoted with 
filled circles. Upper and right panels report the orbital (except for J0801 and 
J0958) and spin period distributions of the confirmed systems including 
the sample presented in this work. Confirmed IPs are taken from 
\cite{ferrario15,bernardini15,esposito15,cotizelati16,tomsick16,johnson17}.
{\it Right:} The softness ratio of polars (filled triangles) and of soft IPs (open circles)
versus blackbody temperature. Values for polars are taken from \cite{ramsay04b} and
for IPs from \cite{anzolin08,masetti12,bernardini12,joshi16} and this work.}
\label{fig:spinorb}
\end{center}
\end{figure*}

While IPs were already known to populate the orbital period
distribution above the 2--3 h gap, the new identifications have revealed
systems with longer periods (e.g. above 6 h). At present there are 14 such IPs 
that represent $\sim 10\%$ of the whole CV population in this period range. 
These long period systems have probably entered 
in the CV phase with evolved donors and  their spin-orbit
period ratios (Figure \ref{fig:spinorb}, left panel) suggest they will reach 
synchronism while evolving to short orbital periods. 
Furthermore, 10 IPs are found below the orbital period gap, where they are not expected \citep{norton04,norton08}, 
indicating that they will likely never synchronise. 
They have low X-ray luminosity, with four identified in the
hard X-ray surveys (EX\,Hya, V1025\,Cen, DO\,Dra, and IGR\,J18173-2509) at only
$\rm L_{14-195\,keV} \sim 1-5 \times 10^{31}\,erg\,s^{-1}$. 
J0801 has a similar luminosity and may in future join this group, once
its orbital period is measured. Finding new low-luminosity IPs
is crucial to confirm the suggestion that short-period IPs 
are intrinsically common as the more luminous long-period IPs \citep{pretorius14}. 
Furthermore, seven out of the 11 polars detected in the hard X-ray 
surveys, including the last addition of J0706, are also found at similar 
luminosities. Whether magnetic systems, and particularly IPs, are
major constituents of the Galactic Centre X-ray emission (GCXE) and of GRXE is 
still largely debated. A recent
re-evaluation using both \XMM\ and {\it NuSTAR} observations of the
inner central (few pc) regions of the Galaxy \citep{hailey16} shows that
relatively massive ($\sim$0.9\,M$_{\odot}$) magnetic WDs can 
account for the X-ray luminosity function at 
$\lesssim 5\times 10^{31}$ erg s$^{-1}$. WDs in IPs are 
indeed found to be relatively massive 
\citep{brunschweiger09,yuasa10,bernardini12,bernardini13,tomsick16} 
as is the case for the majority of the sources in our sample 
(where $\rm <M_{WD}>\sim0.88 M_{\odot}$). This is broadly in line  
with the general  finding that WD primaries in CVs are more massive
than single WDs and WDs in pre-CV binaries, in contrast with
predictions of standard CV formation theory. This disagreement led to suggest
that either the WDs grow in mass during CV evolution, or 
a significant fraction of the observed population of CVs has formed through
a preceding short phase of thermal timescale mass transfer (TTMT), during
which the WD substantially increases in mass \citep{zorotovic11}.
However, neither the inclusion of a TTMT phase nor the adjustment of the accretion efficiency
to allow the WD mass growth in binary population synthesis models 
\citep{wijnen15} are able to explain the observed WD mass distribution in CVs, leaving
the issue still unsolved.
Furthermore, other recent studies
may indicate that the GRXE is not composed primarily of IPs, but rather
by non-magnetic CVs and polars \citep {reis13,warwick14,xu16},  with IPs populating
the harder tail. Thus, this issue may eventually be solved with the identification 
of a larger sample of low-luminosity MCVs and IPs in particular.

Our study has also revealed additional systems displaying a soft
X-ray blackbody component with temperatures ranging from those
observed in the polars ($\sim40-60$ eV) as in J1701, to higher
values ($\sim80$ eV) as in J0958, with distinct softness ratios. 
Since the time of the compilation by \cite{anzolin08}
the current roster of "soft" IPs increased and now amounts to 19 systems
(Figure 6, right panel), representing
$\sim30\%$ of the whole subclass \citep{bernardini12,masetti12,joshi16}. 
Although high blackbody temperatures arising from the WD surface would be locally super-Eddington,
the possibility that the soft component originates instead in the coolest
regions of the accretion flow (e.g. above the WD surface) is not supported by the spectral fits. 
On the other hand, J0706 adds to the increasing number of polars discovered without a 
detectable soft component \citep[see e.g.][]{ramsay09,bernardini14,worpel16}.
This soft component was initially considered the defining characteristics of highly magnetic systems, 
but \XMM\ has demonstrated that not to be the case. Whether this component
is in some cases cool and thus simply shifted to the EUV band, will be difficult to
assess due to the lack of foreseen facilities in this energy range.

We also note that the spectra analysed here can simply
be modeled with a one or two temperature emitting plasma 
(although a more physical model is also used), likely because 
we are sampling X-ray faint targets, where the fine details of the spectra 
are hidden in the noise. A detailed mapping of the plasma conditions 
in the post shock region with high resolution X-ray spectroscopy is clearly
needed. It will also help in solving the puzzle of the hot and 
locally super-Eddington blackbody component. However, 
all this will need to await the next 
generation of X-ray telescopes, like {\it Athena}.

\section*{Acknowledgments}

This work is based on observations obtained with \XMM , an ESA science mission 
with instruments and contributions directly funded by ESA Member States; 
with \Swift, a National Aeronautics and Space Administration (NASA) science 
mission with Italian participation; with \INT , an ESA project with instruments
and science data centre funded by ESA member states and the participation 
of Russia and the USA. This work makes use of observations from the Las Cumbres Observatory (LCO). 
This research has made use of the data supplied by the UK \Swift\ Science
Data Centre at the University of Leicester. This work has also made use of the
APASS catalog, located at the AAVSO web 
site. Funding for APASS has been provided by the Robert Martin Ayres Sciences 
Fund. The REM observations were obtained under programme DDT-REM:32905.
The REM team is acknowledged for the support in the scheduling and data delivery.
This publication also makes use of
data products from: the Wide-field Infrared Survey Explorer, which
is a joint project of the University of California, Los Angeles, and
the Jet Propulsion Laboratory/California Institute of Technology,
funded by the National Aeronautics and Space Administration; the
Two Micron All Sky Survey (2MASS), a joint project of the University
of Massachusetts and the Infrared Processing and Analysis
Center (IPAC)/Caltech, funded by NASA and the NSF; and the
Sloan Digital Sky Survey (SDSS).
DdM and FB acknowledge financial support from ASI/INAF under contract I/037/12/0.

\bibliographystyle{mn2e}
\bibliography{biblio}

\vfill\eject
\end{document}